\documentclass[journal]{IEEEtran}
\usepackage{graphicx}
\usepackage{epstopdf}
\usepackage{multirow}
\usepackage{amsmath}
\usepackage{color}
\usepackage{xcolor}
\usepackage{amssymb}
\usepackage{float}
\usepackage{booktabs}
\usepackage{multirow}
\usepackage{ntheorem}
\usepackage{array}

\usepackage[linesnumbered,algoruled,boxed,lined]{algorithm2e}

\theoremseparator{:}
\newskip\theorempreskipamount
\newskip\theorempostskipamount

\theorembodyfont{}

\ifCLASSINFOpdf
\else
\fi
\begin{document}
	
	%
	\title{Joint Scheduling and Trajectory Optimization of Charging UAV in Wireless Rechargeable Sensor Networks}
	%
	%
	%
	
	\author{Yanheng Liu,
		Hongyang Pan,
		Geng Sun, \IEEEmembership{Member,~IEEE,}
		Aimin Wang,
		Jiahui Li, and 
		Shuang Liang
		
		\thanks{This study is supported in part by the National Natural Science Foundation of China (62172186, 62002133, 61872158, 61806083), in part by the Science and Technology Development Plan Project of Jilin Province (20190701019GH, 20190701002GH, 20210101183JC, 20210201072GX), in part by the Young Science and Technology Talent Lift Project of Jilin Province (QT202013), and in part by Graduate Innovation Fund of Jilin University (101832020CX176, 101832020CX177). (\emph{Corresponding author: Geng Sun.})}
		
		\thanks{Yanheng Liu, Hongyang Pan, Geng Sun, Aimin Wang, Jiahui Li and Shuang Liang are with College of Computer Science and Technology, Jilin University, Changchun 130012, China, and also with Key Laboratory of Symbolic Computation and Knowledge Engineering of Ministry of Education, Jilin University, Changchun 130012, China. (e-mail: yhliu@jlu.edu.cn, panhongyang18@foxmail.com, sungeng@jlu.edu.cn, wangam@jlu.edu.cn, lijiahui0803@foxmail.com, liangshuang8587@foxmail.com)}
		\thanks{Copyright (c) 20xx IEEE. Personal use of this material is permitted. However, permission to use this material for any other purposes must be obtained from the IEEE by sending a request to pubs-permissions@ieee.org.}
		
		\thanks{This manuscript has been accepted by IEEE Internet of Things Journal, doi: 10.1109/JIOT.2021.3132015.}}
	
	%
	%

	\markboth{Journal of \LaTeX\ Class Files,~Vol.~14, No.~8, August~2015}%
	{Shell \MakeLowercase{\textit{et al.}}: Bare Demo of IEEEtran.cls for IEEE Journals}
	%



	\maketitle
\begin{abstract}
	Wireless rechargeable sensor networks with a charging unmanned aerial vehicle (CUAV) have the broad application prospects in the power supply of the rechargeable sensor nodes (SNs). However, how to schedule a CUAV and design the trajectory to improve the charging efficiency of the entire system is still a vital problem. In this paper, we formulate a joint-CUAV scheduling and trajectory optimization problem (JSTOP) to simultaneously minimize the hovering points of CUAV, the number of the repeatedly covered SNs and the flying distance of CUAV for charging all SNs. Due to the complexity of JSTOP, it is decomposed into two optimization subproblems that are CUAV scheduling optimization problem (CSOP) and CUAV trajectory optimization problem (CTOP). CSOP is a hybrid optimization problem that consists of the continuous and discrete solution space, and the solution dimension in CSOP is not fixed since it should be changed with the number of hovering points of CUAV. Moreover, CTOP is a completely discrete optimization problem. Thus, we propose a particle swarm optimization (PSO) with a flexible dimension mechanism, a $K$-means operator and a punishment-compensation mechanism (PSOFKP) and a PSO with a discretization factor, a 2-opt operator and a path crossover reduction mechanism (PSOD2P) to solve the converted CSOP and CTOP, respectively. Simulation results evaluate the benefits of PSOFKP and PSOD2P under different scales and settings of the network, and the stability of the proposed algorithms is verified. 
\end{abstract}


%
%
\section{Introduction} 
\label{introduction}

\par Wireless sensor networks (WSNs) are practical techniques that are extensively utilized in Internet of Things (IoT), e.g., smart homes \cite{8933037} \cite{8640127} \cite{8214099}. A WSN is composed of many sensor nodes (SNs) that are randomly deployed in certain areas, and once deployed, the locations of these SNs are usually fixed and the network can be used to conduct monitoring and communication \cite{2008Wireless} tasks. In order to ensure that SNs work safely and effectively, it is necessary to ensure the power supply demand of SNs \cite{7362107}. However, SNs are usually deployed in unavailable areas, which means that it is impractical to replace the batteries frequently \cite{2020Improving}. Moreover, a large number of discarded batteries will lead to the environmental pollutions. Thus, many approaches are dedicated to optimizing the network protocols \cite{6750328} to prolong the lifetime of WSNs. Although these strategies can prolong the lifetime of WSNs \cite{8737403} to some extent, the limited battery capacity is still a challenge \cite{8060523} for a single SN. 
	
\par With the rapid development of wireless power transfer (WPT) technology \cite{9491798} \cite{9155274} \cite{2019Minimizing}, wireless rechargeable sensor networks (WRSNs) that combine WSNs and WPT can largely solve the battery capacity problem \cite{6515984}. Since the number of SNs in a WRSN is usually large, it is not economical to use a fixed charger for replenishing energy to each node \cite{7470933} \cite{7842599}. Thus, using mobile devices such as unmanned aerial vehicles (UAVs) instead of fixed chargers becomes a feasible and widely used solution \cite{ju2021edge} \cite{2018mTS}. However, due to the limited energy of UAVs and the demand for charging timeliness, it is necessary to reasonably schedule UAVs to improve the charging efficiency.

\par The scheduling of charging UAV (CUAV) is a key point to improve the charging efficiency between a CUAV and SNs in WRSN since it is better to charge all SNs in a shorter time. However, a CUAV may face several challenges for achieving the abovementioned purpose. For example, in a large-scale WRSN with large number of SNs, the number of CUAV hovering points will be huge, causing the increasing of CUAV hovering energy consumption and the reduction of CUAV endurance time. Moreover, some of SNs may be repeatedly charged, such that affecting the total charging efficiency. In addition, the flight order and trajectory of CUAV needs to be considered since it will directly affect the flying overhead of CUAV.
	
\par To solve the abovementioned challenges, we aim to jointly consider the hovering points and flying distance of CUAV for improving the charging efficiency. The main contributions of this work are summarized as follows:
	
\begin{itemize}
	\item Different from the previous works that usually aim to charge SNs by using the static charging piles \cite{2018Charging} or mobile charging vehicles \cite{7924317}, we propose to use a CUAV to charge all SNs in a WRSN so that improving the charging performance of the network. Specifically, we formulate a joint-CUAV scheduling and trajectory optimization problem (JSTOP) to simultaneously minimize the hovering points of CUAV, the number of the repeatedly covered SNs and the flying distance of CUAV for charging all SNs. 
		
	\item JSTOP is proven as NP-hard, and it has both continuous and discrete solution spaces. Moreover, it cannot be directly solved by using conventional algorithm in one step since solving the third objective function needs to determine the hovering locations of CUAV in advance, which needs to be determined by solving the first and second objective functions. Thus, we divide the original formulated JSTOP into two sub-problems that are CUAV scheduling optimization problem (CSOP) and CUAV trajectory optimization problem (CTOP), respectively. 
		
	\item CSOP is also a hybrid optimization problem that consists of the continuous and discrete solution space, and the solution dimension in CSOP is not fixed since it should be changed with the number of hovering points of CUAV. Moreover, CTOP is a completely discrete optimization problem. Thus, we propose two different improved versions of particle swarm optimization (PSO) algorithms, i.e., PSO with a flexible dimension mechanism, a $K$-means operator and a punishment-compensation mechanism (PSOFKP) and PSO with a discretization factor, a 2-opt operator and a path crossover reduction mechanism (PSOD2P), to solve converted CSOP and CTOP, respectively. The proposed charging approach as well as the algorithms can ensure that all SNs are effectively charged in the shortest time so that achieving the maximum energy utilization efficiency and charging efficiency of CUAV.
	
	\item Simulations are conducted to evaluate the effectiveness and performance of the proposed CUAV-based charging approach. Specifically, we show the benefits of the proposed PSOFKP and PSOD2P for solving the formulated CSOP and CTOP under different scales and settings of the network. Moreover, the stability of the proposed algorithms is verified.  
\end{itemize}

\par The remainder of this paper is organized as follows. Section \ref{related work} reviews the related work. Section \ref{system model} shows the system model. Section \ref{problem formulation} formulates JSTOP and gives the corresponding analysis. Section \ref{Algorithmsection} proposes the algorithms. Section \ref{Simulaton results} shows the simulation results. Finally, the conclusion is given in Section \ref{Conclusion}.
%
%
\section{Related work} 
\label{related work}

\par Many previous works have considered installing antennas on UAVs that can transmit RF waves from UAVs to SNs previously \cite{8931359} \cite{9249591} \cite{8530629}. However, as a potential technology of WPT, ER technology is used more and more in real scenarios \cite{8968585}. The former usually needs to satisfy a high requirement of air attitude of UAVs. In other words, once the transmitting antennas on UAVs are not directed at the receiving antenna on the ground, the transmission efficiency is greatly vulnerable. Instead, the latter is robust in tolerating a mite deviation from the predetermined angle \cite{2018Power}. Thus, this work focuses on the application of ER technology to UAV charging, and we introduce a summary of some related work to show the novelty of our work.
	
\par Several previous works study the problem of the coverage of the UAVs. For example, a tethered UAV coverage plan by building a UAV networking system is proposed in \cite{9148486}, and the authors design the network architecture and UAV platform of tethered UAV and proves that the plan can satisfy the requirements of the emergency communication. The authors in \cite{8485481} propose a game-theoretic framework to perform coverage maximization and power control which both of which have Nash equilibrium points, then multi-UAV energy-efficient coverage deployment algorithm based on spatial adaptive play is adopted to guarantee optimal energy-efficient coverage deployment. Liu et al. \cite{9066133} use the reinforcement learning method to build a novel UAV coverage model, which makes a UAV consume less energy by reducing UAV action times when covering the same size area. Moreover, they propose a reliable method to solve the huge problem of the dimension space of the value function. The authors in \cite{8888747} propose a novel antenna array with parasitic elements and install it on UAVs. The results show that the method can improve the coverage range by approximately $156\%$. However, none of these works above considers the scheduling and trajectory of UAV. 

\par Some previous works optimize the deployment and scheduling of CUAVs in WRSNs to enhance the charging performance. For instance, the authors in \cite{8968585} propose a method to maximize the energy efficiency of a CUAV by adjusting its trajectory, and the proposed method has advantages in charging performance and running time. However, SNs may be charged repeatedly, which may cause unnecessary resource consumption. Najeeb et al. \cite{8594255} propose a practical algorithm that can charge SNs without knowing the power level in advance. However, the flight energy consumption of CUAVs is not considered. Xu et al. \cite{9066043} adopt a Q-learning model in which the agent constantly explores and optimizes the scheduling strategy of a UAV, and the method improves the stability of the system by a maximum of $78\%$ compared with no charging scheme. However, the flight freedom of CUAVs is limited, which is impractical for the real scenarios. Chen et al. \cite{8887242} use WPT technology to charge a CUAV, and the deployment of CUAV and the power of the charging station are optimized such that maximizing the overall power transmission efficiency. However, SNs are not considered. Yang et al. \cite{9345142} propose a controllable and obstacle avoidance CUAV routing algorithm to minimize the number of dead SNs in the network and maximize the energy efficiency during charging process. However, the problem of repeated charging is not considered. Zhu et al. \cite{8641332} investigate how to deploy a rechargeable directional sensor network using a mobile charger with the least number of nodes for perpetual target coverage subject to the limited sensing angles of directional sensors and limited energy capacity of the mobile charger. However, the motion energy of the mobile charger is not considered. Li et al. \cite{2020Improving} use the joint deployment of multiple CUAVs to improve the number of SNs in the charging range, maximize the minimum value of charging efficiency and reduce the movement energy consumption of CUAVs. However, the trajectories of CUAVs are ignored. Accordingly, these works above only consider to optimize the positions of CUAVs but do not jointly change the positions and amount of hovering points. Moreover, the number of repeatedly covered SNs is also not considered.
	
\par Moreover, after the scheduling of CUAV, it is necessary to determine the trajectory to further reduce the flight energy consumption of CUAV. There are several existing works that consider the trajectory optimization of UAVs. For example, Salamat et al. \cite{8902345} propose a trajectory generation algorithm to represent the trajectory of UAVs and form a tool to estimate the test location and state. However, they do not consider any background about communication or charging. Ding et al. \cite{9354739} optimize the UAV trajectory based on graph theory in order to minimize the mission completion time of UAV and ensure the connectivity requirements simultaneously. Compared to the conventional method, the performance of this new method is similar, while the complexity is greatly reduced. However, how to properly distribute the mission to separate the flight areas of UAVs is still a problem. Ebrahimi et al. \cite{8960453} adopt a novel reinforcement learning method to make UAV find its trajectory autonomously so as to improve the positioning accuracy of multiple objects and reduce energy consumption of UAV. However, the stability of the proposed method is ignored. Yh et al. \cite{2021Trajectory} design a method that optimizes the flight path of UAV and adaptively controls the interference power on the suspicious link to enhance the average surveillance rate in a limited flight time. However, the algorithm only perform advantages while solving small scale problem. The authors \cite{2018Iterative} boost the security of a relay assisted wireless communication system by optimizing the flight trajectory of UAV. However, they ignore the complete coverage of SNs in WRSN. Wu et al. \cite{9206056} propose a novel evolutionary algorithm to achieve wireless power and information transfer with SNs, then jointly maximize the energy utilization efficiency of UAVs, and minimize the communication delay by optimizing the trajectory. However, increasing the number of UAVs may cause increase the overhead.

\par In recent years, there are several previous works that consider to use the evolutionary algorithms, e.g., PSO and its variants, to solve the different optimization problems \cite{8741295} \cite{9498989}. For example, Zhang et al. \cite{2018Dual} present a novel PSO which can deal with the irregular and unpredictable noise in real-life applications. Although the algorithm has obvious advantages in most test functions, it is not adaptive to solve the discrete problem. Shao et al. \cite{2013Hybrid} introduce a multi-objective hybrid PSO for solving flexible job-shop scheduling problem and this algorithm considers the discrete solution space. However, it ignores the hybrid solution condition which consists of both continuous and discrete solution space. Liang et al. \cite{2006Comprehensive} use a novel learning strategy whereby historical best information of all other particles is used to update the velocity of the particle. This method enables the diversity of the swarm to be preserved to discourage premature convergence. However, it does not consider a priori information, which may cause additional unnecessary cost. Moreover, Cao et al. \cite{2019Comprehensive} embed the local search strategy in conventional PSO for achieving higher convergence rate and accuracy. However, it may make the algorithm fall into local optima in some conditions. The authors \cite{2005Application} propose a parallel PSO with a special operator to obtain faster optimization. However, the improvement of algorithm performance is limited.

\color{black}{
%
%
\section{System models}
\label{system model}

\par In this section, the network model, wireless charging model and energy consumption model of CUAV are introduced.

%
%
\subsection{Network model}
\begin{figure}
	\centering{\includegraphics[width=3in]{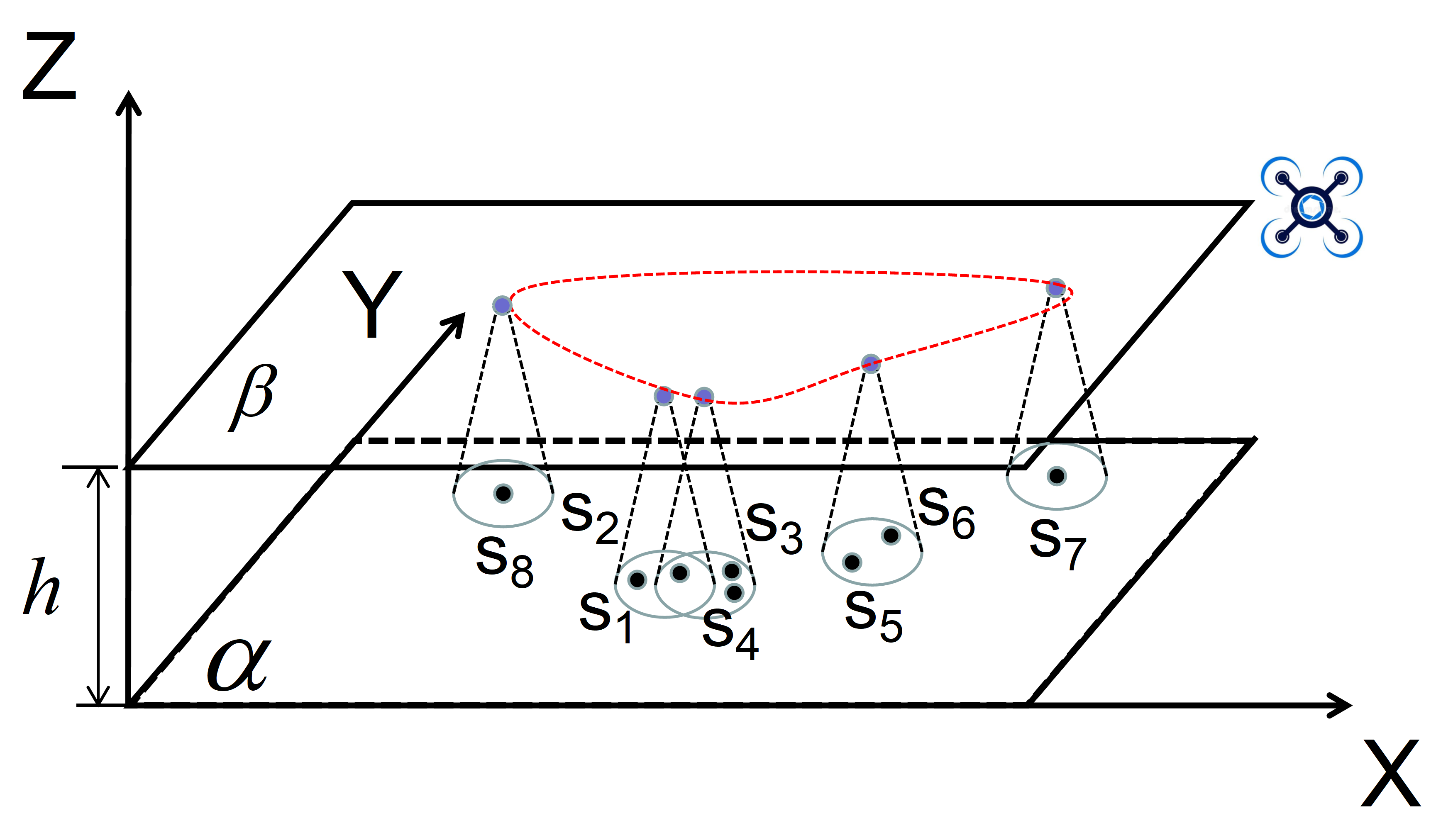}}
	\caption{System model of a CUAV-enabled WRSN.}
	\label{System_model}
\end{figure}

\par As shown in Fig. \ref{System_model}, we consider a CUAV-enabled WRSN system that consists of a CUAV and $n$ SNs recorded as $S_i(i\in\{1,~2,~3,~...,~n\})$. The flight height of CUAV for charging SNs is fixed at $h$, which means that CUAV flying at the plane $\beta$. Moreover, SNs are randomly deployed on the plane $\alpha$, and each of them can be powered by a CUAV. CUAV visits different hovering positions and it can simultaneously charge several SNs within the effective charging range such that all nodes can get power eventually. The 3D coordinates of SNs are defined as $S_i=(x_i, y_i, 0)$, and thus, the distance $D_{us}$ between CUAV $q(t)=(x(t), y(t), h)$ and $S_i$ can be achieved as follows:

\begin{equation}
\begin{aligned}
	D_{us}&={||q(t)-S_i||}_2\\
	&=\sqrt{(x(t)-x_i)^2+(y(t)-y_i)^2+h^2}
\end{aligned} 
\end{equation}
	
\noindent where $ 0 \leq t \leq T$, and $T$ represents the total flight time of CUAV.

%
%
\subsection{Wireless charging model}

\par The far-field directional radio frequency (RF) wave \cite{8639104} and electromagnetic resonance (ER) \cite{9161953} are the most commonly used WPT technologies. Specifically, the former has a long transmission distance, while it has the strict requirements on the positions of transmitters. If the transmitting antenna and receiving antenna are not aligned in an RF system, the transmission efficiency of the system may be greatly affected. However, the transmission efficiency of the latter is less affected by the alignment and the transmission efficiency can reach more than $90\%$ \cite{8968585}. Thus, ER is regarded as a promising technology, and it is adopted in this work to supply energy from CUAV to SNs. Specifically, the charging efficiency $\eta$ of the ER technology is defined as follows \cite{8968585}:

\begin{equation}
\begin{aligned}
	\eta=\frac{a}{b{{D}_{us}}^6+c}
\end{aligned}
\end{equation}

\noindent where $a$, $b$ and $c$ are constants that are related to the parameters of CUAV. Refer to \cite{8968585}, we set $a=b=c=1$ in this work for simplicity. Thus, $D_{us}$ is the only variable, which is variated according to the positions of CUAV and SNs, and the specific variation range of $D_{us}$ will be given in Section \ref{Simulaton results}.
	
%
%
\subsection{Energy consumption model of CUAV}

\par In this paper, a multi-rotor aero-robot is adopted due to its hover function in the sky. Wu et al. \cite{8968585} extends the energy consumption model of this type of CUAV into a CUAV-enabled WRSN. Specifically, the energy consumption of CUAV can be divided into propulsion energy consumption and charging energy consumption. We only need to consider the horizon propulsion energy consumption of the propulsion energy consumption because we assume that CUAV flies at the same altitude in this work. It is worth noting that the energy consumption caused by CUAV acceleration is usually ignored \cite{8968585}. Accordingly, the propulsion energy consumption model of CUAV can be expressed as follows\cite{2018Energy}:
	
\begin{equation}
\label{E_pro}
\begin{aligned}
	E_{pro}(T)=\int_{0}^{T}P_{pro}(||v(t)||)dt
\end{aligned}
\end{equation}
	
\noindent where $T$ is the flight time, $v(t)$ is the speed of CUAV, and $P_{pro}$ is the propulsion power which can be calculated as follows:
	
\begin{equation}
\label{P_pro}
	\begin{aligned}
		P_{pro}(v)=&P_0(1+\frac{3v^2}{U_{tip}^2})+P_i{(\sqrt{1+\frac{4v^4}{4{v_0}^4}}-\frac{v^2}{2{v_0}^2})}^{1/2}\\&+\frac{1}{2}d_0\rho sA{v}^3
	\end{aligned}
\end{equation}
	
\noindent where $P_0$ and $P_i$ are the two constant parameters for the blade profile power and induced power at the hovering status, respectively. $U_{tip}$ is the tip speed of the rotor blade, $v_0$ is the mean rotor induced velocity when hovering, $d_0$ and $s$ represent the fuselage drag ratio and rotor solidity, respectively, and $\rho$ and $A$ are the air density and rotor disc area, respectively. Assume that CUAV flight speed is fixed, then the propulsion power of CUAV} can be divided into the moving power and hovering power, respectively. The moving power $P_{move}$ is expressed as follows:   
	
\begin{equation}
	\begin{aligned}
		P_{move}=P_{pro}(v(t)=v_{move})
	\end{aligned}
\end{equation}
	
\noindent where $v_{move}$ is the moving speed of CUAV. Moreover, the hovering power $P_{hover}$ is expressed as follows:
	
\begin{equation}
	\begin{aligned}
		P_{hover}=P_{pro}(v(t)=0)
	\end{aligned}
\end{equation}
	
\noindent If we assume that the total CUAV movement time and the total CUAV hovering time are $t_m$ and $t_h$, respectively, then the total energy consumption of CUAV can be rewritten as follows:
	
\begin{equation}
\label{E_total}
	\begin{aligned}
		E_{total}(t_m,t_h)=P_{move}t_m+(P_{hover}+P_s)t_h
	\end{aligned}
\end{equation}

\noindent where $P_s$ represents the transmission power from CUAV to SNs.

%
%
\section{Problem statement}
\label{problem formulation}

\par In this section, JSTOP for improving the performance of the WRSN system is formulated and analyzed.
	
%
%
\subsection{Problem formulation}
\par In this work, we propose to use a CUAV to charge all SNs in WRSN. Specifically, CUAV flies at the fixed altitude and it is assumed the time for charging each SN is the same due to low energy storage capacity of each SN. Moreover, CUAV needs to hover at a point for charging a set of SNs within its effective charging range and then moves to the next position one by one until all SNs are charged, which can extend the lifetime of WRSN. Note that a CUAV can charge multi SNs at each hovering point if these SNs are within its effective charging range. However, there may be three challenges for CUAV in the abovementioned charging process. \textbf{First}, the number of the hovering points of CUAV will be huge, especially for a large-scale network with large number of SNs, which may increase the hovering energy consumption of CUAV so that further affecting the endurance time of CUAV. \textbf{Second}, some of SNs may be repeatedly charged, as shown in Fig. \ref{System_model}, which may waste some charging power so that decreasing the total charging efficiency. \textbf{Finally}, CUAV needs to fly over every prior determined hovering point to charging all SNs, and thus how to select flight order and trajectory will directly affect the flying distance as well as the propulsion energy consumption of CUAV.
	
\par Accordingly, in the considered scenario, the ultimate objective is to improve the charging performance between CUAV and SNs in the UAV-enabled WRSN. In other words, we aim to let CUAV charge all SNs with the least hovering points and the shortest flying distance. Note that improving the overall charging performance can be directly affected by three optimization objectives as follows.

\par \textbf{\emph{Optimization objective 1: Minimizing the number of CUAV hovering points.}} According to Eq. (\ref{E_total}), the total hovering energy consumption of CUAV is linear with the total hovering time of CUAV (i.e., the total charging time) since the charging time for each SN is assumed to be the same. Thus, CUAV should select the less hovering points while covering more SNs at each hovering point so that reducing the hovering energy and time consumptions of CUAV, thereby improving the charging efficiency to a certain extent. Note that the vital constraint of this optimization objective is that the determined hovering points of CUAV should cover all SNs, i.e., each SN must be charged in each hovering point of CUAV. Therefore, we think that it is necessary to optimize the number of CUAV hovering points, and the corresponding objective function is designed as follows:   
	
\begin{equation}
	\label{f1}
	\begin{aligned}
		f_1(X_{u1},~Y_{u1},...,~X_{uk},~Y_{uk},~k)= k
	\end{aligned}
\end{equation}
	
\noindent where $(X_{ui},~Y_{ui}), i=1,~2,~3,~...,~k$ is the two-dimensional coordinates of CUAV.

\par \textbf{\emph{Optimization objective 2: Minimizing the number of repeatedly covered SNs.}} As mentioned above, the number of SNs that within the charging range of CUAV at each hovering point should be maximized to improve the charging efficiency. However, this may lead some SNs to be repeatedly charged, especially when SNs are densely distributed, which may reduce the total charging efficiency. Thus, it is reasonable to reduce the number of repeatedly covered SNs in each hovering point of CUAV, and the corresponding objective function is designed as follows:

\begin{equation}
	\label{f2}
	\begin{aligned}
		f_2(X_{u1},~Y_{u1},...,~X_{uk},~Y_{uk},~k)= S_{rc}\\	
	\end{aligned}
\end{equation}
	
\noindent where $S_{rc}$ represents the number of repeatedly covered SNs by the different locations of CUAV, and it is expressed as follows: 

\begin{equation}
  \label{S_{rc}}
     \begin{aligned}
       S_{rc}=\sum_{j=1}^{k}\sum_{i=1}^{n}{u}_{ij}-n\\
     \end{aligned}
\end{equation}

\noindent where ${u}_{ij}$ is a binary variable, which is designed as follows: 

\begin{equation}
\label{u_{ij}}
	\begin{aligned}
		{u}_{ij}=\left\{\begin{matrix}
		1& & {||q_j-S_i|| \leq d_{max}}\\
		0& & {||q_j-S_i|| > d_{max}}
	\end{matrix}\right.  
\end{aligned}
\end{equation}
\noindent where $q_j$ is the $j$th hovering location.

\par \textbf{\emph{Optimization objective 3: Minimizing the flying distance of CUAV for charging all SNs.}} After determining the optimal hovering points for charging, CUAV needs to fly over each prior determined hovering point to charging all SNs. Thus, how to select the flight order and plan the trajectory becomes a vital problem to reduce the total flying distance of CUAV, which directly affects the total charging performance, and the corresponding objective function is designed as follows:
	
\begin{equation}
\label{f3}
	\begin{aligned}
		f_3=\sum_{a,b=1}^{k}c_{ab}d_{ab}	
	\end{aligned}
\end{equation}

\noindent where $d_{ab}$ is the distance between the hovering locations $q_a$ and $q_b$, wherein $a=1,~2,~3,~...,~k$ and $b=1,~2,~3,~...,~k$. Moreover, $c_{ab}$ is expressed as follows:

\begin{equation}
  \label{c}
    \begin{aligned}
      c_{ab}=\left\{
        \begin{matrix}
           1,&~\text{If CUAV visits}~q_b~\text{after visiting}~q_a\\ 
           0,&~\text{otherwise}
        \end{matrix}\right.
\end{aligned}
\end{equation}

\par Accordingly, the abovementioned three optimization objectives are all critical to the charging performance of UAV-enabled WRSN, and they are jointly considered in this work. Thus, JSTOP can be formulated as follows:

\begin{subequations}
	\label{JSTOP}
	\begin{flalign}
		 \text{(JSTOP)}
		 \min \limits_{\{X_{u}, Y_{u}\}, \{q_j\}, k}& f=k+S_{rc}+\sum_{a,b=1}^{k}c_{ab}d_{ab}\\
		  \text{s.t.}\qquad &X_{umin}\leq X_{u}\leq X_{umax} \\
		  \qquad&Y_{umin}\leq Y_{u}\leq Y_{umax}\\
		  \qquad&k\in \mathbb{N}^+, 2\leq k\leq n\\
		  \qquad&n_{c}= n\\
		  \qquad &c_{ab}\in \{0, 1\}\\
		  \qquad &\sum_{a=1}^{k} c_{ab}=1\\
		  \qquad &\sum_{b=1}^{k} c_{ab}=1	 
	\end{flalign}
\end{subequations}

\noindent where $n_{c}$ represents the total number of SNs that can obtain power and $S$ represents the set of SNs.

%
%
\subsection{Problem conversion}

\par The formulated JSTOP is NP-hard, and it has both continuous and discrete solution spaces, which is difficult to be solved by using conventional algorithms. Moreover, JSTOP cannot be directly solved since solving the third objective function needs to determine the hovering locations of CUAV in advance, which needs to be determined by solving the first and second objective functions. Thus, we convert the original formulated JSTOP into two sub-problems that are CSOP and CTOP, respectively. Specifically, CSOP is formulated as follows:
	
\begin{subequations}
	\label{CSOP}
	\begin{align}
		\text{(CSOP)}\quad
		\min \limits_{\{X_{u}, Y_{u}\}, k} &f_{CSOP}=f_{1}+f_{2}\\
		\text{s.t.}\quad &X_{umin}\leq X_{u}\leq X_{umax} \\
		\qquad&Y_{umin}\leq Y_{u}\leq Y_{umax}\\
		\qquad&n_{c}= n\\
		\qquad&k\in \mathbb{N}^+, 2\leq k\leq n 
	\end{align}
\end{subequations}
	
\noindent Then, CTOP is formulated as follows:

\begin{subequations}
\label{CTOP}
	\begin{align}
		\text{(CTOP)}\qquad
		\min \limits_{\{q_j\}}\quad & f_{CTOP}=f_{3}\qquad\\
		\text{s.t.}\quad &c_{ab}\in \{0, 1\}\\
		\qquad &\sum_{a=1}^{k} c_{ab}=1\\
		\qquad &\sum_{b=1}^{k} c_{ab}=1	 	
	\end{align}
\end{subequations}

\par By the conversions above, the original formulated JSTOP can be solved by using different algorithms, which will be detailed in the following section.

%
%
\subsection{Problem analysis}
\par In this section, the NP-hardness of the converted CSOP and CTOP are analyzed.

%
%
\par For CSOP, as shown in Eq. (\ref{CSOP}), it is constructed by the linear weighting method of $f_1$ and $f_2$. For ease of analysis and without loss of generality, we only consider $f_1$ and thus the original formulated CSOP is simplified as follows: 

\begin{subequations}
	\label{CSOP-SIM}
	\begin{align}
	\min \limits_{\{X_{u}, Y_{u}\}, k} &f_1=k\\
	\text{s.t.} \quad&n_{c}= n
	\end{align}
\end{subequations}

\noindent \textbf{Definition 1} (Minimum geometric unit disk cover problem (MGUDCP)): Assume that a set of points $P={p_1, p_2, ..., p_n}$ and a set $C$ on the plane, and the goal is that ${\cup}_{{C}_i \in C}{C}_i$ covers all points $P$. Then, the goal is to select the minimum cardinality subset ${C}^{\prime} \subseteq C$ such that each of the points in $P$ is covered by at least one disk in ${C}^{\prime}$. 

\par Thus, the optimization problem shown in Eq. (\ref{CSOP-SIM}) above can be reduced to MGUDCP \cite{2012On}, which is an NP-hard problem and it is transformed from 3SATm which is a classical NP-hard problem \cite{1987The}. Since the formulated CSOP is more complex than the simplified problem in Eq. (\ref{CSOP-SIM}), then the original CSOP is NP-hard. 

%
%
\par For CTOP, as shown in the Eq. (\ref{CTOP}), it can be reduced to a traveling salesman problem (TSP), which is NP-hard \cite{8067863}. Thus, CTOP is also an NP-hard problem. 

\par Accordingly, the original formulated JSTOP is an NP-hard problem since both of CSOP and CTOP are NP-hard.

%
%
\section{Algorithms}
\label{Algorithmsection}

\par The formulated optimization problem is proven as NP-hard, which means that an algorithm cannot find the optimal solution in polynomial time. Therefore, one effective strategy is to use the evolutionary algorithms to find a feasible solution in finite iterations, which is a practical method for dealing with the considered problem. Evolutionary algorithms have many advantages in solving NP-hard problems \cite{2020Joint}. However, the conventional PSO cannot satisfy the solving requirements of CSOP and CTOP, since the discrete optimization is included in these solutions. Therefore, we propose two different version of improved PSO algorithms for dealing with these two converted optimization problems, respectively. 

%
%
\subsection{Conventional PSO}

\par PSO is originally proposed to solve the continuous optimization problems \cite{488968}. Compared to other evolutionary algorithms, PSO has several advantages, e.g., fewer parameters and ease of use \cite{2013A}. Moreover, PSO has a memory function, which means that the best history position of the particle population can be recorded, and these solutions can share the information with other solutions in the population. In addition, its nature of interacting particles has proved to be effective in finding the optimum in the search space. Thus, these characteristics motivated us to choose PSO as the basic algorithm structure for solving the formulated optimization problems.

\par In conventional PSO, each member of the group constantly changes its velocity and position by learning from its own and other members' experiences. Let $\nu_d^m=(\nu_1^m, \nu_2^m, ..., \nu_{D}^m)$ and ${\chi}_d^m=({\chi}_1^m, {\chi}_2^m, ..., {\chi}_{D}^m)$ be the velocities and positions of the $m$th individual, respectively, where $m=1, 2, 3, ..., U$ and $U$ represents the population size, and suppose that ${P_{best_{d}}}^m=({P_{best_{1}}}^m, {P_{best_{2}}}^m, ..., {P_{best_{D}}}^m)$ and $G_{best_d}=(G_{best_1}, G_{best_2}, ..., G_{best_D})$ denote the local optimal solution and the global optimal solution, respectively, where $d=1, 2, 3, ..., D$ and $D$ is the dimension of decision variables. Then, the population update process of conventional PSO can be expressed as follows \cite{488968}: 

\begin{subequations}
	\label{popupdate}
	\begin{align}
	    \nu ^m_d=&w*\nu^m_d+c_1*r_1*({P_{best_{d}}}^m-{\chi}^m_d) \notag \\
     	&+c_2*r_2*(G_{best_d}-{\chi}^m_d)\\
     	{\chi}_d^m=&\nu^m_d+{\chi}_d^m
	\end{align}
\end{subequations}

\noindent where $w$ is the inertia weight, $c_1$ and $c_2$ are the learning factors, and $r_1$ and $r_2$ denote two random numbers produced in the range from $(0, 1)$. The main steps of conventional PSO can be found in \cite{488968}.

%
%
\subsection{Motivation for proposing the two PSO variants}

\par As mentioned in Section IV, the formulated JSTOP is NP-hard, and it has both continuous and discrete solution spaces, which is difficult to be solved by using conventional PSO. Moreover, JSTOP cannot be directly solved by using conventional PSO in one step since solving the third objective function needs to determine the hovering locations of CUAV in advance, which needs to be determined by solving the first and second objective functions. Thus, in this work, the original formulated JSTOP are converted into CSOP and CTOP, respectively. However, there are still several challenges for conventional PSO to deal with the converted two sub-problems. 

\begin{itemize}
	\item CSOP is also a hybrid optimization problem that consists of the continuous and discrete solution space, which becomes a challenging task for conventional PSO since it is proposed for the continuous optimization problems. 
	
	\item The solution dimension in CSOP is not fixed since it should be changed with the number of hovering points of CUAV, and this cannot be handled by conventional PSO. 
	
	\item CTOP is a completely discrete optimization problem, which means that conventional PSO cannot be used to solve it.
	
	\item The positions of CUAV, the flying path of CUAV and the number of hovering points of CUAV are the solution to the formulated optimization problem. Thus, we must optimize (2 $\times$ $k$+1) solution dimensions. Moreover, the number of solution dimension increases in proportion to both of the network scale and number of SNs, which means that the converted CSOP may become a large-scale optimization problem. For example, if we have 1000 SNs in a WRSN, the upper limit of the dimension of the solution can reach 2001 at most, which is difficult to be solved.

\end{itemize}

\par Accordingly, we tried to seek for some previous works that related to PSO for finding out the suitable advanced PSO variants to solve the formulated optimization problems, which are mainly presented in Section \ref{related work}. However, there is no ready-made method that can be directly used to solve the converted problems, which motives us to propose PSOFKP and PSOD2P, and they are detailed as follows.	

%
%
\subsection{PSOFKP}

\par CSOP is actually a hybrid optimization problem since it has both continuous and discrete solution spaces, which means that it cannot be directly solved by using conventional PSO. Thus, we propose a PSOFKP with three improved factors to solve the formulated CSOP, and the details are as follows.

%
%
\subsubsection{Flexible dimension mechanism}

\par In CSOP, the number of CUAV hovering points and the corresponding hovering positions are regarded as the solutions. However, the former is not preset, which means that the dimension of a solution cannot be determined. Thus, we introduce a flexible dimension mechanism to dealing with the dimension length change condition, which is detailed as follows. 

\par
\begin{figure*}[ht]
	\centering{\includegraphics[width=7in]{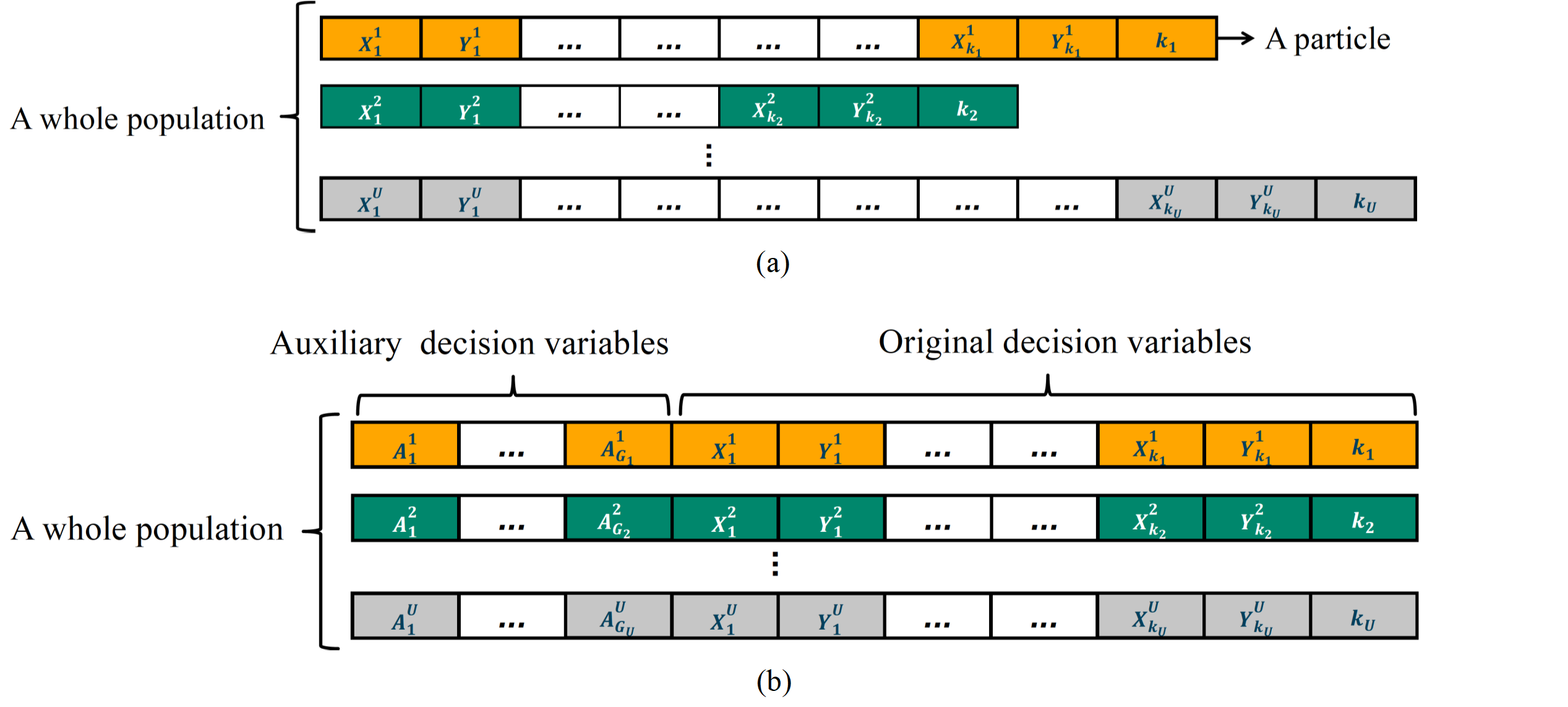}}
	\caption{(a) The entire population after the flexible dimension mechanism. (b) The entire population with auxiliary variables.}
	\label{syn dimension}
\end{figure*}

 Fig. \ref{syn dimension}(a) shows the basic principle of the flexible dimension mechanism. Specifically, this mechanism adds the auxiliary variables to the particles with nonmaximum dimensions to make all the particles reach the maximum dimension. These auxiliary variables are randomly generated in the search space to ensure that the algorithm works properly and do not participate in the calculation of the objective function. The description of the auxiliary variables production process is shown in Fig. \ref{syn dimension}(b).

%
%
\subsubsection{$K$-means operator}

\par In conventional PSO, the initial solutions are usually randomly generated. However, a high quality heuristic function can improve the search efficiency and performance of the algorithm \cite{8527444}. Moreover, we find that when using the update process of the particle position shown in Eq. (\ref{popupdate}), the convergence becomes weaker as the iteration progress. Thus, the $K$-means operator that considers the locations of SNs is applied to initialize the solutions and the iteration process in the algorithm. Specifically, suppose that $k$ is an integer that generated randomly from $[2, n]$ and the decision variable can be written as ($X_{u1}, Y_{u1}, ..., X_{uk}, Y_{uk}, k$), wherein $n$ is the number of SNs. Moreover, the $K$-means clustering is used to determine $g$ centers from a group of points after minimizing the mean squared distance between each given point and the nearest center. The main steps of the $K$-means operator are described in \cite{1017616}.

\par In this work, the locations of SNs play roles as the given points and $k$ is given as the number of centers. After applying the $K$-means operator, we can obtain a set of hovering positions, and the process is shown in Algorithm \ref{Kmeans}. Thus, the initialization process may be more efficient which can improve the performance of the algorithm, and the algorithm can obtain a better convergence rate.

\begin{algorithm}
	\caption{$K$-means operator for determining the hovering positions of CUAV}
	\label{Kmeans}
	Obtain the coordinates of SNs;\\
	Taking the sensor coordinates as the given points, the coordinates of $k$ cluster centers are obtained according to \cite{1017616};\\
	The coordinates of $k$ cluster centers are injected into the population.
\end{algorithm}

%
%
\subsubsection{Punishment-compensation mechanism}

\par Due to the uncertainty of $k$, the algorithm may easily become trapped in the local optima, causing it to be unable to guarantee that all SNs are charged. Thus, we propose a punishment-compensation mechanism to overcome the shortcomings above. Specifically, the value of $k$ is changed iteratively, and the process follows the two following principles:

\textbf{(a) Punishment}: If $k$ of the local optimal solution of the particle is greater than that of the global optimal solution, $k$ decreases by $\frac{n}{100}$. Otherwise, $k$ of the global optimal solution is assigned to the particle.

\textbf{(b) Compensation}: If $k$ of the local optimal solution of the particle is less than that of the global optimal solution, $k$ increases by $\frac{n}{100}$. Otherwise, $k$ of the global optimal solution is assigned to the particle. Moreover, in order to ensure that the search step size is not too large, we provide a usage probability $\rho$ for this mechanism. 

\par The main steps of the proposed punishment-compensation mechanism are presented in Algorithm \ref{PC}. Note that $rand$ is a random number that generated from $[0, 1]$, $k^i$ is the $k$ value of particle $i$, $k^{P_{best_i}}$ is the $k$ value of the local optimal solution of the particle $i$ and $k^{G_{best}}$ is the $k$ value of the global optimal solution.

\begin{algorithm}
	\caption{Punishment-compensation mechanism}
	\label{PC}
	\textbf{Set the related parameters $\rho$;}\\
		\eIf{all SNs can obtain energy}{
			\If{$rand<\rho$}{
				\eIf{$k^{P_{best_i}}\ge k^{G_{best}}$}{$k^i=k^i-n/100;$}{$k^i=k^{G_{best}};$}
				\If{$k^i\le 1$}{$k^i=k^i+n/100;$}	
			}
		}{
			\If{$rand<\rho$}{
				\eIf{$k^{P_{best_i}}\le k^{G_{best}}$}{$k^i=k^i+n/100;$}{$k^i=k^{G_{best}};$}
				\If{$k^i\ge n$}{$k^i=k^i-n/100;$}
			}
		}	
\end{algorithm}

\par Accordingly, proposed PSOFKP is described in Algorithm \ref{PSOFKP}.
	
\begin{algorithm}
	\caption{PSOFKP}
	\label{PSOFKP}
	\textbf{Define the fitness function:}
	$f_{CSOP}(x), x=[x^1,~x^2,...,~x^{U}];$\\
	\textbf{Set the related parameters.}\\
	(1) The learning factors: $c_1$ and $c_2$;\\
	(2) The inertia weight: $w$;\\
	(3) The maximum number of iterations: $G_{max}$;
	\For{m=1 to U}
	{
		Initialize the velocity $\nu^m$ and position ${\chi}^m$ with Algorithm \ref{Kmeans};\\
		Compute $f_{CSOP}({\chi}^m)$ and set ${P_{best}}^m={\chi}^m$;	
	}
	$G_{best}=\min ({P_{best}}^m)$;\\
	\For{it=1 to $G_{max}$}
	{
		\For{m=1 to U}
		{	
			Update the value of $k$ with Algorithm \ref{PC};\\
			Generate additional population position $\Lambda^m$ according to the newly generated $k$ with Algorithm \ref{Kmeans};\\
			Compute $f_{CSOP}(\Lambda^m)$;\\
			Update the velocity $\nu^m$ and position ${\chi}^m$ according to Eq. (\ref{popupdate});\\
			Compute $f_{CSOP}({\chi}^m)$;\\
		}
		In order to maintain the population size, the redundant particles were eliminated at position $\Lambda$ and position ${\chi};$\\
		\For{m=1 to U}
		{
			\If{$f_{CSOP}({\chi}^m)<f_{CSOP}({P_{best}}^m)$}{${P_{best}}^m={\chi}^m$;}
			\If{$f_{CSOP}({P_{best}}^m)<f_{CSOP}(G_{best})$}{$G_{best}={P_{best}}^m$;}
		}
	}
	Output $G_{best}$.\\
\end{algorithm}

\par It is worth noting that the ``$<$'' in Algorithm \ref{PSOFKP} is different from the general meaning. Specifically, it judges whether the two results meet the same prerequisite, that is, whether they can charge all nodes or not, and then compares the corresponding values. If one result can charge all nodes and the other can not, then even if the value of the former is greater than that of the latter, the former is still less than the latter.

%
%
\subsection{PSOD2P}

\par Although several versions of discrete PSO algorithms have been proposed to solve the discrete optimization problems, they may be not suitable to be directly used for the formulated CTOP, and the reasons are discussed in Section V-B. Thus, a PSOD2P with three improved factors is proposed and the details are as follows.

%
%
\subsubsection{Discretization improvement factor}

\par We redefine the velocity and position, and the calculation method between them in conventional PSO as follows. 
	
\noindent \textbf{Definition~2} (Positions and velocity): Let the access sequence of UAV hovering locations be the solution of CTOP, i.e. the position in PSOD2P. The exchange pair $W(A,~B)$ is defined as the exchange of two hovering positions $A$ and $B$ in the solution sequence, and the ordered queue of the set of exchange pairs is the velocity. For example, assume that the solution sequence is $Q=q_1,~q_2, ... ,~q_i,~q_j,~q_t,~...,~q_k$ where $k$ is the total amount of hovering. Then, $Q^*=Q+W(q_i,~q_j)=q_1,~q_2,...,~q_j,~q_i,~q_t,~...,~q_k$. Assume that the velocity is $L=({W}_1(q_i,~q_j),~{W}_2(q_j,~q_t)))$, then  $Q^*=Q+L=q_1,~q_2,...,~q_t,~q_i,~q_j,~...,~q_k$. 
	
\noindent \textbf{Definition~3} (Equivalent velocity): Different velocities acting on the same solution may produce the same new solution. These velocities are called equivalent velocity.
	
\noindent \textbf{Definition~4} (Velocity merging): Several velocities can be combined to obtain a new velocity, and $\textcircled{+}$ is defined as the merging operator of two velocities.
	
\noindent \textbf{Definition~5} (Simplest velocity): Among all equivalent velocities, the velocity with the fewer exchange pairs is called the simplest velocity. For example, $Q^*-Q=W(q_i,~q_j)$ is a simplest velocity.
	
\par Overall, Eq. (\ref{popupdate}) can be rewritten as follows:
	
\begin{subequations}
	\label{popupdate2}
	\begin{align}
	\nu^{m\prime}=&w*\nu^{m\prime}\textcircled{+}r_1*({P_{best}}^{m\prime}-{\chi}^{m\prime}) \notag \\
	&\textcircled{+}r_2*(G_{best}-{\chi}^{m\prime})\\
	{\chi}^{m\prime}=&\nu^{m\prime}+{\chi}^{m\prime}
	\end{align}
\end{subequations}

\noindent where $w$ is a constant, $r_1$ and $r_2$ are two random numbers that are generated from $[0, 1]$, $w*\nu$ represents all exchange pairs in simplest velocity $\nu$ reserved by probability $w$ and $r_1*({P_{best}}^{m\prime}-{\chi}^{m\prime})$ represents all exchange pairs in simplest velocity $({P_{best}}^{m\prime}-{\chi}^{m\prime})$ reserved by probability $r_1$. In the same way, $r_2*(G_{best}-{\chi}^{m\prime})$ represents all exchange pairs in simplest velocity $(G_{best}-{\chi}^{m\prime})$ reserved by probability $r_2$. Obviously, the larger the parameters $w, r_1$ and $r_2$ are set, the more exchange pairs that are retained in $\nu, ({P_{best}}^{m\prime}-{\chi}^{m\prime})$ and $(G_{best}-{\chi}^{m\prime})$, respectively.

%
%
\subsubsection{2-opt operator} 

\par
\begin{figure}
	\centering{\includegraphics[width=3in]{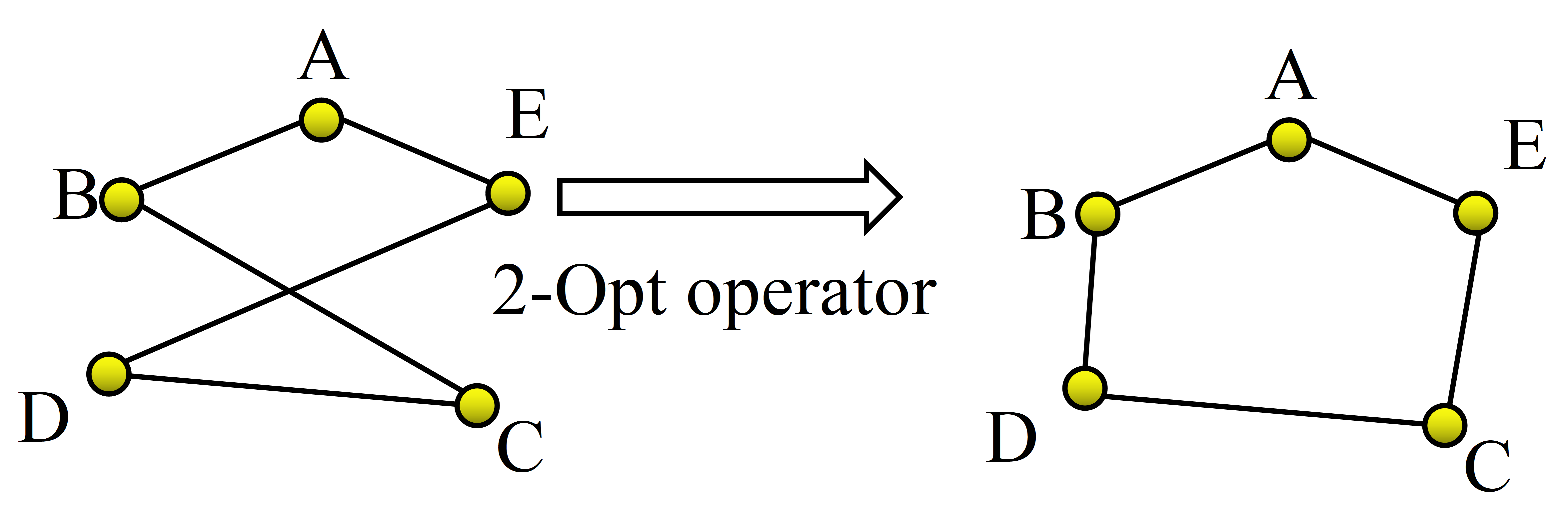}}
	\caption{The access sequence before and after the 2-opt process.}
	\label{2-opt}
\end{figure}

 The 2-opt algorithm is a classic local search method that can accelerate the convergence of the algorithm and improve the accuracy of the solution, and it is widely used in the graph optimization problems \cite{2017A}. The idea of the 2-opt operator is to remove two edges from a route and reconnect these edges in order to obtain a new route with a shorter total path length. Fig. \ref{2-opt} shows the basic principle of 2-opt operator. Specifically, the original access sequence is $A, B, C, D, E, A$, and after the 2-opt process, we reconnect the edge $B, C$ and $D, E$. Then, the access sequence is changed to $A, B, D, C, E, A$. Moreover, this removal and reconnection process should be continued until no 2-opt improvement is found. The process of 2-opt operator is shown in Algorithm \ref{2OPT}, and the details are described as follows.

\begin{algorithm}
\caption{2-opt operator}
\label{2OPT}
	\For{j=1 to n-3}{
		\For{h=j+2 to n-1}{
			\If{$E[B_{j,j+1}]+E[B_{h,h+1}]>E[B_{j,h}]+E[B_{j+1,h+1}]$}{
				Swap the access sequence of $B_{j+1}$ and $B_h$;
			}
		}	
	}

\end{algorithm}

\par In the algorithm, $B$ is a legitimate access sequence, and $B_{j+1}$ and $B_h$ represent the ${(j+1)}_{th}$ and $h_{th}$ hovering positions of CUAV, respectively. Moreover, $E[B_{j,j+1}]$ is the Euclidean distance between $B_j$ and $B_{j+1}$, $E[B_{h,h+1}]$ is the Euclidean distance between $B_h$ and $B_{h+1}$, $E[B_{j,h}]$ is the Euclidean distance between $B_j$ and $B_{h}$, and $E[B_{j+1,h+1}]$ is the Euclidean distance between $B_{j+1}$ and $B_{h+1}$.
	
%
%
\subsubsection{Path crossover reduction mechanism}

\par
\begin{figure}
	\centering{\includegraphics[width=3in]{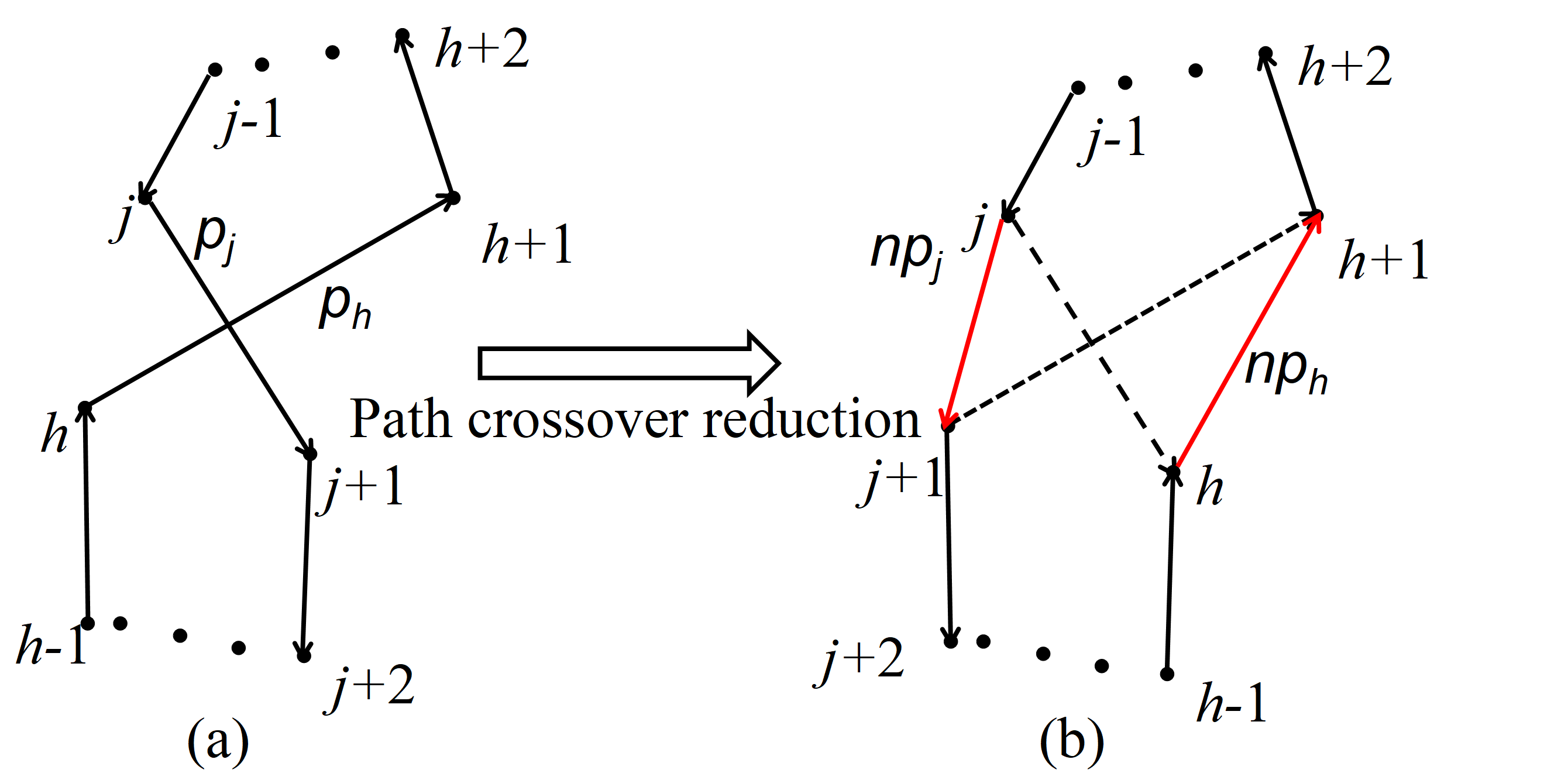}}
	\caption{(a) The origin path before the using of path crossover reduction mechanism. (b) The path after using the path crossover reduction mechanism.}
	\label{crossreduce}
\end{figure}

 Using 2-opt operator can produce new crossover paths. Besides, it may also slow down the convergence rate of the algorithm. Thus, in this section, we introduce a path crossover reduction mechanism to solve the issues above. Specifically, path ${p}_j$ is defined as the path composed of vertex $j$ and vertex $j+1$, which is shown in Fig. \ref{crossreduce}(a). Similarly, path ${p}_h$ is defined as the path composed of vertex $h$ and vertex $h+1$, and there is an intersection between ${p}_j$ and ${p}_h$, and the trajectory of CUAV optimized by reducing the crossover is shown in Fig. \ref{crossreduce}(b). According to the principle that the sum length of two sides of a triangle in a plane is greater than the third side length, it is easy to prove ${p}_j+{p}_h>{np}_j+{np}_h$. Moreover, we define a function $iscro$ whose function is to determine whether two line paths in a plane is intersected. If they do, 1 is returned; otherwise, 0 is returned. Then, the path crossover reduction mechanism can be described is Algorithm \ref{pathreduce}.
	
\begin{algorithm}
	\caption{Path crossover reduction mechanism}
	\label{pathreduce}
	\For{j=1 to n-2}{
		$h=j+2$;\\
		\While{h=j+2$\leq$n}{
			\If{iscro(${p}_j,{p}_h$)}{
				Reverse the sequence from node $j+1$ to node $h$;\\
				$h=j+2$;
			}
			$h=h+1$;
		}	
	}
\end{algorithm}

\par In the algorithm, $n$ is the amount of hovering CUAV. Then, PSOD2P can be described in Algorithm \ref{PSODCTOPOPT}.

\begin{algorithm}[t]
	\caption{PSOD2P}
	\label{PSODCTOPOPT}
	\textbf{Define the fitness function:}
	$f_{CTOP}(x), x=[x^1,~x^2,...,~x^{U}];$\\
	\textbf{Set the related parameters.}\\
	(1) The learning factors: $c_1$ and $c_2$;\\
	(2) The inertia weight: $w$;\\
	(3) The maximum number of iterations: $G_{max}$;
	\For{m=1 to U}
	{
		Initialize the velocity $\nu^{m\prime}$ and position ${\chi}^{m\prime}$;\\
		Compute $f_{CTOP}({\chi}^{m\prime})$ and set ${P_{best}}^{m\prime}={\chi}^{m\prime}$;	
	}
	${G_{best}}^\prime=\min ({P_{best}}^{m\prime})$;\\
	\For{it=1 to $G_{max}$}
	{
		\For{m=1 to U}
		{	
			Compute the simplest velocity $({P_{best}}^{m\prime}-x^m)$ and ${G_{best}}^\prime-x^m)$;\\
			Compute the velocity $\nu^{m\prime}$ according to Eq. (\ref{popupdate2}a);\\
			Compute the position ${\chi}^{m\prime}$ according to Eq. (\ref{popupdate2}b) and compute $f_{CTOP}({\chi}^{m\prime})$;\\
			\If{(mod(it,20)==0)}
			{Compute the new position ${\chi}^{m\prime\prime}$ according to Algorithm \ref{2OPT};\\
			\If{$f_{CTOP}({\chi}^{m\prime\prime})<f_{CTOP}({\chi}^{m\prime})$}{${\chi}^{m\prime}={\chi}^{m\prime\prime}$}}
			\If{(mod(it,20)==0)}
			{Update the position ${\chi}^{m\prime}$ according to Algorithm \ref{pathreduce};\\
			\If{$f_{CTOP}({\chi}^{m\prime})<f_{CTOP}({P_{best}}^{m\prime})$}{${P_{best}}^m={\chi}^m$}
			\If{$f_{CTOP}({P_{best}}^{m\prime})<f_{CTOP}(G_{best})$}{${G_{best}}^\prime={P_{best}}^{m\prime}$}}	
			Update the velocity $\nu^{m\prime}$ to an simplest velocity;
		}
	}
	Output ${G_{best}}^\prime$.\\
\end{algorithm}

%
%
\subsection{Convergence and complexity of the proposed algorithms} 

\par In this section, the convergence and complexity of the proposed PSOFKP and PSOD2P are analyzed.

%
%
\subsubsection{Convergences of the proposed algorithms}

\par Evolutionary algorithms are effective methods to solve the NP-hard problems since they are able to find a feasible solution under the limited time and resources without considering the specific form of the problem. Specifically, according to the principle, the evolutionary algorithms can find a stable suboptimal solution by utilizing continuous iterations. Moreover, the authors in \cite{2008Convergent} analyze the convergence of the evolutionary algorithms and give two lemmas on the probability of locating in the promising area and the different distance from the initial solution to the optimal solution, which proves that the evolutionary algorithm will be sure to be convergent for any tolerance whatever for unimodal or multimodal functions. Moreover, the convergence of the proposed algorithms can be further observed and verified by the simulation results in the next section.

%
%
\subsubsection{Complexities of the proposed algorithms}

\par For PSOFKP, we assume that the maximum iteration, the number of SNs, the number of CUAV hovering points and the population size are $t$, $n$, $k$ and $U$, respectively, then the computation of $K$-means has $O(t \cdot n \cdot k \cdot I)$ complexity, as mentioned in \cite{9514914}, where $I$ is the processing time for calculating the distance between two hovering positions. Note that $I$ can be ignored with the increasing of the computing power, hence the computation complexity of $K$-means can be regarded as $O(t \cdot n \cdot k)$. Due to the uncertainty of $k$ of each solution, the worst overall computation complexity of PSOFKP is $O(U \cdot t \cdot n^{2})$.

\par For PSOD2P, the complexity of PSOD2P is primarily derived by the comparison of the 2-opt operator and the path crossover reduction mechanism. We assume that the solution dimension and population size are $n$ and $U$, respectively, then the comparison of in 2-opt operator has $O(U \cdot n^{2})$ computational complexity. Moreover, the path crossover reduction mechanism also has $O(U \cdot n^{2})$ computational complexity. Thus, the overall complexity of proposed PSOD2P is $O(U \cdot n^{2})$.

%
%
\section{Simulation results}
\label{Simulaton results}

\par In this section, simulations are conducted to evaluate the performance of the proposed approaches by using Matlab. We consider three simulation cases with different numbers of SNs (100 SNs for Case 1, 500 SNs for Case 2 and 1000 SNs for Case 3), and the region of WRSN is set as 500 m $\times$ 500 m. Moreover, the flight height $h$ is 10 m, and the maximum transmission distance $d_{max}$ between CUAV and SNs is $10 \sqrt{2}$ m. In addition, the population size and the maximum number of iterations of PSOFKP and PSOD2P are 20 and 200, respectively. Note that the simulations are repeated independently for 30 times to avoid the random bias, and the statistical optimization results are presented.

%
%
\subsection{Optimization of CSOP} 
\label{SubCSOP}

\par
\begin{figure*}[htb]
	\centering{\includegraphics[width=7in]{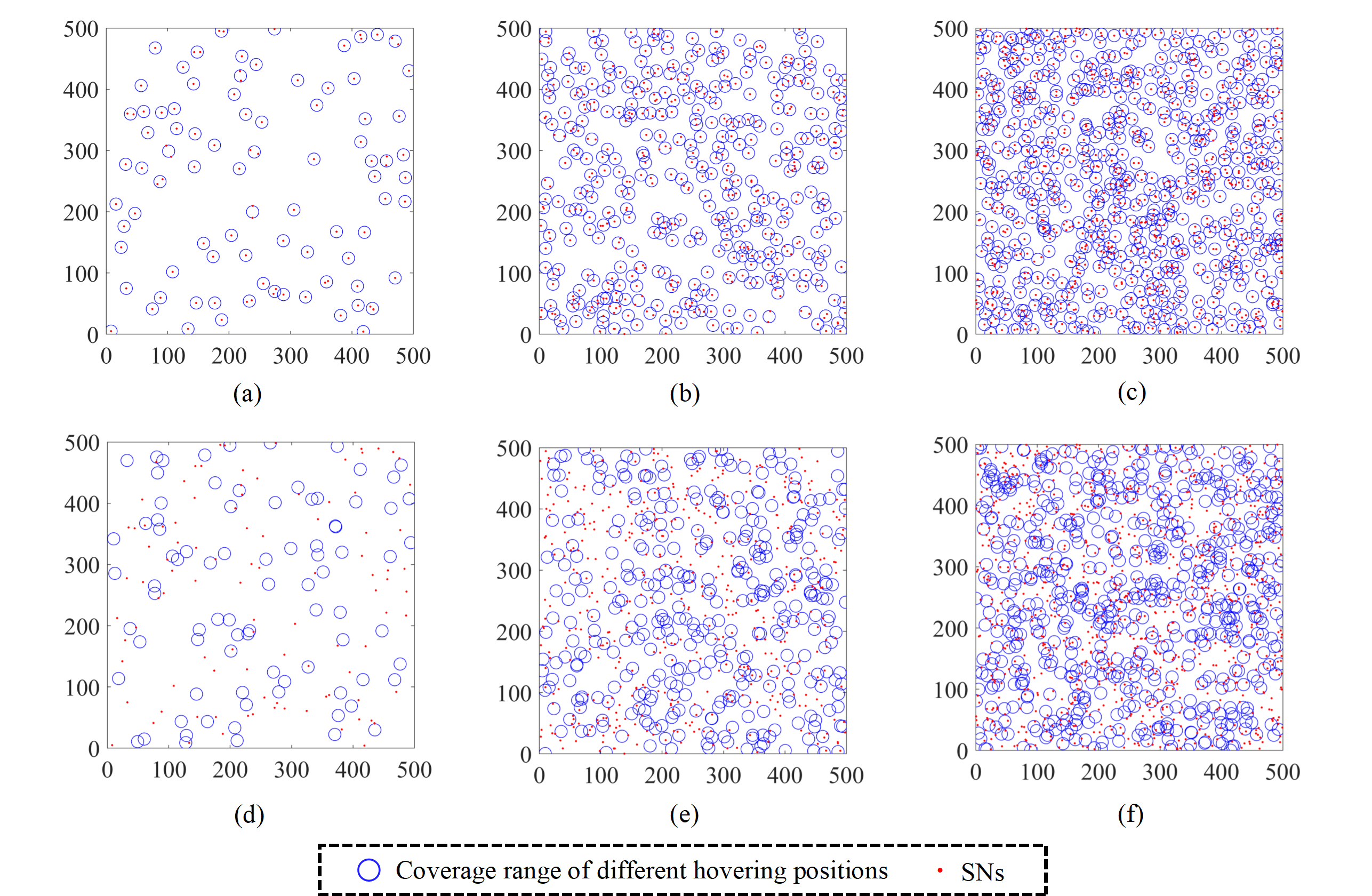}}
	\caption{The optimization results of CUAV hovering positions obtained by PSOFKP and random scheduling. (a) PSOFKP for case 1. (b) PSOFKP for case 2. (c) PSOFKP for case 3. (d) Random scheduling for case 1. (e) Random scheduling for case 2. (f) Random scheduling for case 3.}
	\label{merge}
\end{figure*}

 Figs. \ref{merge}(a), \ref{merge}(b) and \ref{merge}(c) show the scheduling results of CUAV for charging the all SNs obtained by the proposed PSOFKP in different simulation cases, while Figs. \ref{merge}(d), \ref{merge}(e) and \ref{merge}(f) give the random scheduling results of CUAV. Note that the random scheduling means that the hovering points of CUAV are randomly determined. As can be seen, the random scheduling method cannot satisfy the constraint that all nodes should be charged.

\par Then, we introduce the firefly algorithm (FA) \cite{2020Improving} and cuckoo search (CS) \cite{9130775} to solve CSOP for comparison, and use the uniform scheduling method as the comparison approach. The parameters of the abovementioned comparison algorithms are given in Table \ref{tab:parameters1}. Note that the uniform scheduling method means that the number of CUAV hovering points is fixed as 1301, and the hovering points are uniformly distributed in the area. Table \ref{numerical results of CSOP} shows the corresponding numerical results of the proposed PSOFKP and the abovementioned comparison approaches. Note that ``Mean'', ``Std.'', ``Maximum'' and ``Minimum'' represent the mean value, standard deviation, maximum value and minimum value of the 30 independent tests, respectively. Moreover, the statistical results of the running time of FA, CS and the proposed PSOFKP are shown in Table \ref{1numerical results of CSOP}. It can be seen from these tables that PSOFKP achieves the better results compared to other algorithms, while it takes the longer CPU time. This is because that the introduced improved factors of PSOFKP will undoubtedly takes some extra calculations, which may consume more time. However, the gap of running time between proposed PSOFKP and other algorithms is not very large, and the proposed CUAV-based charging method is usually performed off-line. Thus, we may say that PSOFKP has the overall best performance for dealing with converted CSOP.

\begin{table}
	\begin{center}
		\newcommand{\tabincell}[2]{\begin{tabular}{@{}#1@{}}#2\end{tabular}}
		\caption{Parameter setups of different algorithms}
		\label{tab:parameters1}
		\begin{tabular}{l|c}
			\hline
			\textbf{Algorithm}   &\textbf{Values of parameters} \\ 
			\hline
			\multirow{1}{*}{\textbf{FA} }	
			& $\alpha =0.5$, $\beta=0.2$, $\gamma=1$ \\ 			
			\multirow{1}{*}{\textbf{CS}  }
			& $\beta=3$  \\ 
			\multirow{1}{*}{\textbf{PSOFKP}  }	
			& \tabincell{c}{ $c_{1}=2$, $c_{2}=2$, $w=0.73$ , $kt=20$ }\\ 	
			\hline
		\end{tabular}	
	\end{center}
\end{table}

\begin{table}
	\begin{center}
		\caption{Numerical statistical results obtained by PSOFKP and other contrast algorithms for solving CSOP}
		\setlength{\tabcolsep}{1.6mm}
		{\begin{tabular}{l|l|l|l|l|l|l}\hline
				& & &{Mean} &{Std.} &{Maximum} &{Minimum} \\
				\hline
				\multirow{8}{*}{\textbf {Case 1}}
				&\multirow{2}{*}{\textbf {Uniform}}
				&$f_1$  &1301.00 	&0.00	&1301.00 	&1301.00 \\
				&&$f_2$  &61.00	&0.00	&0.00	&0.00	\\
				&\multirow{2}{*}{\textbf {FA}}
				&$f_1$  &93.00 	&3.07	&99.00 	&87.00 \\
				&&$f_2$  &5.27	&3.67	&15.00	&0.00	\\
				&\multirow{2}{*}{\textbf {CS}}
				&$f_1$  &92.87 	&2.54	&98.00 	&88.00 \\
				&&$f_2$  &4.63	&3.10	&13.00	&0.00	\\
				&\multirow{2}{*}{\textbf {PSOFKP}}
				&$f_1$  &\textbf{87.43} 	&1.72	&91.00 	&84.00 \\
				&&$f_2$  &\textbf{0.00}	&0.00	&0.00	&0.00	\\\hline
				\multirow{8}{*}{\textbf {Case 2}}
				&\multirow{2}{*}{\textbf {Uniform}}
				&$f_1$  &1301.00 	&0.00	&1301.00 	&1301.00 \\
				&&$f_2$  &286.00	&0.00	&0.00	&0.00	\\
				&\multirow{2}{*}{\textbf {FA}}
				&$f_1$  &428.07 &23.90	&471.00 	&387.00 \\
				&&$f_2$ &134.70 &40.62	&220.00	    &77.00	\\
				&\multirow{2}{*}{\textbf {CS}}
				&$f_1$  &435.13 	&21.18	&476.00 	&382.00 \\
				&&$f_2$  &143.80	&36.43	&230.00	    &72.00	\\
				&\multirow{2}{*}{\textbf {PSOFKP}}
				&$f_1$  &\textbf{381.97} 	&9.82	&408.00 	&366.00 \\
				&&$f_2$  &\textbf{67.93}	&8.96	&93.00	&45.00	\\\hline
				\multirow{8}{*}{\textbf {Case 3}}
				&\multirow{2}{*}{\textbf {Uniform}}
				&$f_1$  &1301.00 	&0.00	&1301.00 	&1301.00 \\
				&&$f_2$  &579.00	&0.00	&0.00	&0.00	\\
				&\multirow{2}{*}{\textbf {FA}}
				&$f_1$  &787.10 	&57.61	&892.00 	&703.00  \\
				&&$f_2$  &659.30	&149.37	&950.00	&449.00	\\
				&\multirow{2}{*}{\textbf {CS}}
				&$f_1$  &789.47 	&41.72	&895.00 	&708.00 \\
				&&$f_2$  &664.67	&108.55	&949.00	&538.00	\\
				&\multirow{2}{*}{\textbf {PSOFKP}}
				&$f_1$  &\textbf{674.23} 	&15.19	&696.00 	&643.00 \\
				&&$f_2$  &\textbf{389.73}	&24.71	&428.00	&346.00	\\\hline
		\end{tabular}}
		\label{numerical results of CSOP}
	\end{center}
\end{table}

\begin{table}
	\begin{center}
		\caption{Numerical statistical results of CPU times (s) obtained by different algorithms for solving CSOP}
		\setlength{\tabcolsep}{1.6mm}
		{\begin{tabular}{l|l|l|l|l|l}\hline
				& &{Mean} &{Std.} &{Maximum} &{Minimum} \\\hline
				\multirow{3}{*}{\textbf {Case 1}}
				&\textbf{FA} 	&\textbf{4.05}  &0.33 	&5.343  &3.58  \\
				&\textbf{CS} 	&12.82 &0.84  &14.81  &11.44  \\
				&\textbf{PSOFKP} 	&92.35   &19.91  	&114.59  &31.67  \\\hline
				\multirow{3}{*}{\textbf {Case 2}}
				&\textbf{FA} 	&\textbf{36.93}  &0.10 	&39.38  &34.78   \\
				&\textbf{CS} 	&193.31 &9.66  &219.42  &175.13  \\
				&\textbf{PSOFKP} 	&608.37  &147.97 	&798.91 &120.72 \\\hline
				\multirow{3}{*}{\textbf {Case 3}}
				&\textbf{FA} 	&\textbf{193.94}  &2.93 	&201.31   &188.70   \\
				&\textbf{CS} 	&996.28  &75.34  &1148.53  &896.39   \\
				&\textbf{PSOFKP} 	&1630.90   &158.88 	&1988.51  &1340.73  \\\hline
				
		\end{tabular}}
		\label{1numerical results of CSOP}
	\end{center}
\end{table}

\par
\begin{figure*}[htb]
	\centering{\includegraphics[width=7in]{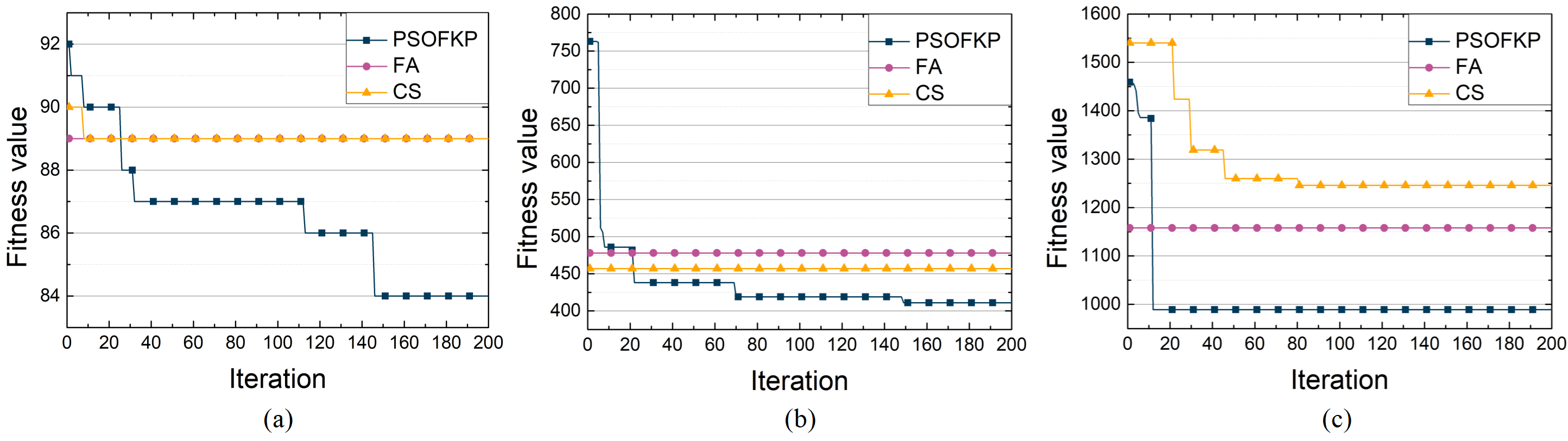}}
	\caption{Convergence rates obtained by PSOFKP. (a) Case 1. (b) Case 2. (c) Case 3.}
	\label{CSOP-convergence}
\end{figure*}

 Fig. \ref{CSOP-convergence} shows the convergence rates of PSOFKP, FA and CS. As can be seen, PSOFKP has the fastest convergence rate compared to the benchmark algorithms. The reason may be that the introduced $K$-means method is able to enhance the performance of conventional PSO. Specifically, this operator is embedded into the initialization process of conventional PSO to make the initial solutions to be more reasonably distributed by utilizing their prior knowledge. Moreover, this operator is also embedded into the solution update process to improve the exploitation ability of the algorithm. In addition, the punishment-compensation mechanism is helpful to update the discrete part of the solutions and enhance the diversity of the population. Thus, these operators may accelerate the convergence rate of the algorithm.

%
%
\subsection{Optimization of CTOP} 

\par In this section, the comparison algorithms for solving the formulated CTOP are introduced in detail, and the optimization results are presented.
%
%
\subsubsection{Introduction to comparison algorithms} 

\par The artificial bee colony (ABC) \cite{8067863} \cite{8556456}, genetic algorithm (GA) \cite{4666932} \cite{9498989}, differential evolution (DE) \cite{0An} \cite{2019Global}, simulated annealing (SA) \cite{9197634} \cite{2011Improved} and tabu search (TS) \cite{5209064} \cite{8627949} are introduced as the comparison algorithms with PSOD2P for solving the converted CTOP. Specifically, how are these abovementioned algorithms applied to solve the investigated problems are presented to make a clearer comparison. 

\par All of these algorithms are evolutionary algorithms, and thus they have the similar algorithm structure. In CTOP, the solutions are the trajectory of CUAV for approaching each hovering point, which means that the solution space is with discrete values. Note that conventional ABC, GA, DE, SA and TS may solve the discrete optimization problems by using the mapping mechanism, and the details are as follows.

\par \textbf{\emph{ABC:}} ABC algorithm is inspired by the foraging behavior of bees. In ABC algorithm, the colony of artificial bees is divided into three groups: employed bees, onlookers and scouts. Half of the colony consists of the employed bees, and another half consists of the onlookers. The position of a food source corresponds to a possible solution to the optimization problem, and the nectar amount of each food source represents their quality (fitness) of the associated solution. The number of the employed bees equals to the number of food sources. When a food source has been abandoned by bees, the abandoned employed bee would become a scout, and the main steps of ABC for solving formulated CTOP are as follows. First, ABC needs to generate the initial solutions and the population randomly, and then evaluates the objective function values of these solutions, wherein the initial solutions are the access sequences of CUAV hovering positions. After the initialization, the population of solutions is subject to repeated cycles of the search courses of employed bees, onlookers and scouts. Second, the algorithm enters the employed bee phase. Specifically, for each food source, the algorithm exchanges two access sequences of CUAV hovering positions in the food source randomly to generate new food source and retain the better food source according to the greedy principle. When all employed bees complete the search process, they will share the information about nectar amounts and positions of food sources with onlookers. An onlooker evaluates the nectar information which is owned by all employed bees, and then chooses a food source with a probability which is related to the nectar amount. As in the case of the employed bee, the onlooker can produce a modification on the position in its memory and check the nectar amount of the candidate source. If the nectar amount is more than that of the previous one, the bee would memorize the new position and forgets the previous one. An onlooker chooses a food source completely depending on the probability value associated with the food source ${\tau}_i$, which is calculated as follows: 

\begin{equation}
\label{ABC-EQ}
\begin{aligned}
{\tau}_i={\xi}_i/\sum_{j=1}^{U}{\xi}_j\\
\end{aligned}
\end{equation}

\noindent where ${\xi}_i$ denotes the fitness value of the $i$th solution which is proportional to the nectar amount of the food source in the $i$th position and $U$ denotes the population size. Third, the food source which is abandoned by the bees would be replaced with a new food source that found by the scouts. In ABC algorithm, the foraging behavior is simulated by randomly producing a food source and replacing the abandoned one with a new one. If a position cannot be further improved through a predetermined number of cycles, the food source should be abandoned. The predetermined number of cycles is an important control parameter in ABC algorithm, which is called limit for abandonment. The generation process of new food source can be described as exchanging two access sequences of CUAV hovering positions in the old food source randomly. If the current iteration time is less than the maximum iteration, the algorithm executes from the second step to the third step for a loop.

\par \textbf{\emph{GA:}} A solution to the formulated optimization problem will be regarded as a chromosome of GA, and the main steps of GA for solving formulated CTOP are as follows. First, GA needs to randomly generate the initial solutions and population, and then evaluates the objective function values of these solutions. Similar to ABC, the solutions of GA are regarded as the access sequences of CUAV hovering positions. Second, for each chromosome, the algorithm should execute the follow operators: 
\noindent \textbf{(a) Crossover:} Two parent individuals are selected to crossover by roulette wheel which are recorded as $Z_1$ and $Z_2$, respectively. If the generated random number is less than the crossover probability, the selected gene fragment from $Z_1$ will be retained in the offspring chromosome, and the length of the gene fragment is $s$, where $s=length(Z)/2-1$, and $length(Z)$ represents the length of the chromosome. The starting position of the gene fragment is randomly generated in the first half of the chromosome of the offspring chromosome, and the rest gene of the offspring chromosome is generated according to $Z_2$. Specifically, the offspring chromosome starts from the position after the end of the gene fragment and puts the gene corresponding to $Z_2$ if the gene is not appeared in offspring chromosome until the length of the offspring chromosome is equal to that of the parent chromosomes. It is worth noted that the chromosome is a cycle sequence, which means that the next gene will become the first one if the gene is the last one in its chromosome. 
\noindent \textbf{(b) Mutation:} If the generated random number is less than the crossover probability, the algorithm needs to exchange two genes in chromosome randomly to generate a new chromosome. It is worth noted that every node of the access sequences of CUAV hovering positions is regarded as gene in GA. If the current iteration time is not satisfied with the requirement of the maximum iteration, then the algorithm should execute the crossover and mutation operator for every chromosome iteratively.

\par \textbf{\emph{DE:}} Based on the natural selection idea of survival of the fittest, DE has the characteristics of self-learning and self-adaptive to a certain extent. At the start of the evolution, the perturbations are large since parent populations are far away. As the evolutionary process matures, the population converges to a small region and the perturbations adaptively become small. An individual in DE represents a solution to the formulated optimization problem and the main steps of DE for solving formulated CTOP are as follows. First, DE needs to generate the initial solutions and the population randomly, and then evaluates the objective function values of these solutions. However, the solution of conventional DE is the real numbers in search space in the initialization, while the solution to the formulated optimization problem is the access sequence of the hovering positions. Thus we transform the real numbers to the sequence by sorting in order to calculate the objective function. The order of magnitude of real numbers in an individual is the access order of CUAV. Second, do mutation, crossover, selection as conventional DE recurrently until the current iteration reaches the maximum iteration. It is worth noted that we compute the objective function values by the method that transform the real numbers to the sequence by sorting. The main steps of DE is shown in \cite{0An}. 

\par \textbf{\emph{SA:}} A solution to the formulated optimization problem will be regarded as a particle of SA, and the main steps of SA for solving CTOP are as follows. First, SA needs to generate $U$ initial solution, set the search times in every temperature as $L$ and the cooling rate as $q$, then evaluates the objective function value, wherein $U$ is the population size. The method of generating the new solutions in SA is to swap the access order of two CUAV hovering positions. Then, for each individual in the population, SA needs to generate $L$ new solutions according this method. Third, the algorithm calculates the objective function values of the new generated solutions and the corresponding increments between new solutions and initial solution. If the new solutions are better than the current optimal solution, then the new solutions should be retained, or retain the new solution according to the probabilities \cite{2011Improved}. Meanwhile, the temperature reduction process is always accompanied by the algorithm, just like conventional SA.

\par \textbf{\emph{TS:}} TS is a memory-based search strategy that allows the local search process to proceed beyond local optima. This is achieved by allowing the objective function to deteriorate when the current solution is a local optimum, and by keeping track of recent moves or solutions in a so-called tabu list. Whenever the algorithm attempts to move to a solution or to perform a move recorded in the tabu list, the move is banned. This rule prevents cycling and forces other solutions to be explored. The main steps of TS for solving CTOP are as follows. First, the TS algorithm needs to generate one initial solution randomly, set the tabu list to empty set, and give the maximum number of candidate solutions and the tabu length as $U$ and $\Psi$, respectively, where the solution is the access sequence of CUAV hovering positions. Second, for each candidate solution, the algorithm exchanges two access sequences of the initial solution to generate $U$ new candidate solutions. Third, TS updates the current optimal solution, the tabu list and the solution to the next iteration process. To avoid falling into local optimum in the search process, this movements will be recorded in the tabu list. Specifically, the tabu list records the access sequence that has been exchanged and the tabu length that limits the length of exchanging action. If the coming exchanging movement has been recorded in tabu list, the algorithm will choose the optimal movement which has not been record in tabu list. Finally, if the algorithm does not meet the termination condition, execute the second and the third step iteratively.

\par Moreover, the key parameter settings of the abovementioned comparison algorithms are given in Table \ref{tab:parameters2}.

\begin{table}
	\begin{center}
		\newcommand{\tabincell}[2]{\begin{tabular}{@{}#1@{}}#2\end{tabular}}
		\caption{Parameter setups of different algorithms}
		\label{tab:parameters2}
		\begin{tabular}{l|c}
			\hline
			\textbf{Algorithm}   &\textbf{Values of parameters} \\ 
			\hline
			\multirow{1}{*}{\textbf{ABC} }	
			& $Limit = 200$ \\ 		
			\multirow{1}{*}{\textbf{GA}  }
			& $pc = 0.9$, $pm = 0.9$  \\ 		
			\multirow{1}{*}{\textbf{DE} }
			& $F= 0.6$, $CR = 0.5$ \\	
			\multirow{1}{*}{\textbf{SA} }
			& $T_0 = 500$,  $L = 200$, $q = 0.98$  \\  	
			\multirow{1}{*}{\textbf{TS}}				
			& $\psi = 200$ \\ 		
			\multirow{1}{*}{\textbf{PSOD2P}   }		
			& $c_1 = 0.8$, $c_2 = 0.8$, $w = 1$   \\ 
			\hline
		\end{tabular}	
	\end{center}
\end{table}

%
%
\subsubsection{Optimization results of CTOP}
\par
\begin{figure*}[htb]
	\centering{\includegraphics[width=7in]{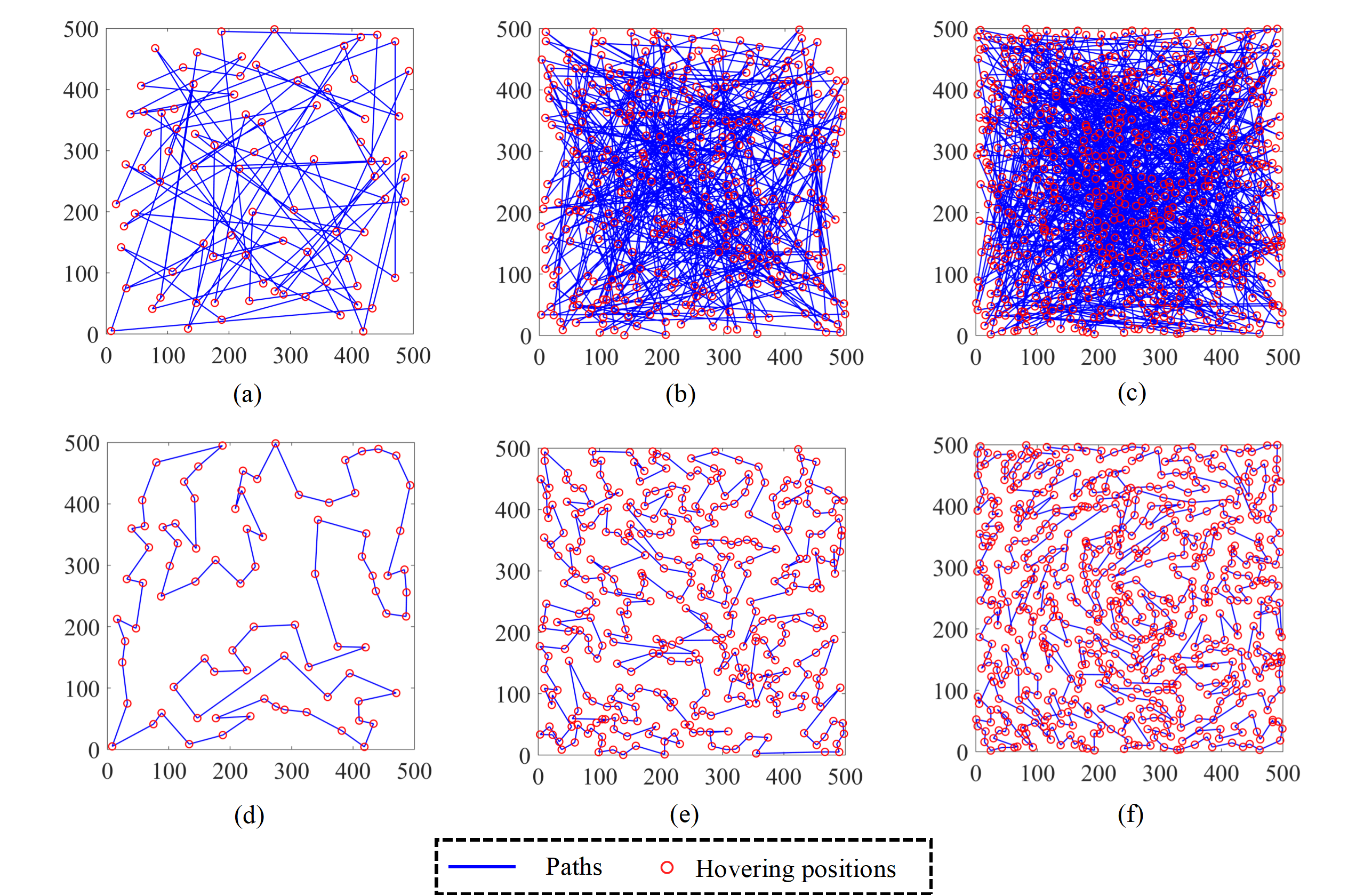}}
	\caption{The original path and the path optimized by PSOD2P. (a) The original path for case 1. (b) The original path for case 2. (c) The original path for case 3. (d) The path optimized by PSOD2P for case 1. (e) The path optimized by PSOD2P for case 2. (f) The path optimized by PSOD2P for case 3.}
	\label{merge2}
\end{figure*}

\begin{figure*}[htb]
	\centering{\includegraphics[width=7in]{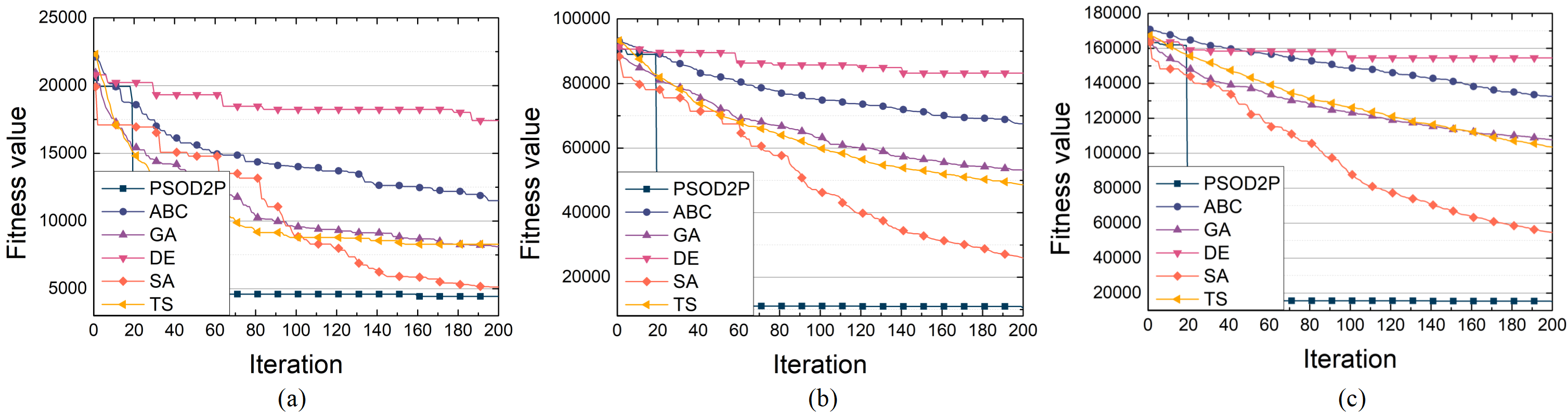}}
	\caption{Convergence rates of different algorithms for solving CTOP in different cases. (a) Case 1. (b) Case 2. (c) Case 3.}
	\label{CTOPconvergence}
\end{figure*}

 Figs. \ref{merge2}(a), \ref{merge2}(b) and \ref{merge2}(c) show the randomly generated  trajectories for CUAV for case 1, case 2 and case 3, respectively, while Figs. \ref{merge2}(d), \ref{merge2}(e) and \ref{merge2}(f) show the trajectories of CUAV optimized by using proposed PSOD2P. Obviously, the flight distance of CUAV is significantly reduced by PSOD2P. Moreover, the numerical statistical results and the running time obtained by different algorithms for solving CTOP are shown in Tables \ref{numerical results of CTOP} and \ref{CPU time results of CTOP}, respectively, and the definitions of the ``Mean'', ``Std.'', ``Maximum'' and ``Minimum'' have the same with meanings with CSOP case. Similar to CSOP case, the running time of PSOD2P increases more than that of other comparison algorithms as the dimension increasing, and the reason may be that PSOD2P has more cyclic search in decision variables. In addition, Fig. \ref{CTOPconvergence} reveals the convergence rates of different algorithms for solving CTOP. As can be seen, the proposed PSOD2P achieves the best performance in terms of the convergence rate and accuracy. 

\begin{table}[h]
	\begin{center}
		\caption{Numerical statistical results obtained by PSOD2P and other comparison algorithms for solving CTOP}
		\setlength{\tabcolsep}{2.2mm}
		{\begin{tabular}{l|l|l|l|l|l}\hline
				
				& &{Mean} &{Std.} &{Maximum} &{Minimum} \\
				\hline
				\multirow{6}{*}{\textbf {Case 1}}
				&\textbf{ABC} 	&12376.32 &425.57	&13181.90&11498.25 \\
				&\textbf{GA}  	&9091.24	&483.78	&10109.88 &8096.96 \\
				&\textbf{DE}  	&18326.82 &440.96	&19244.45&17424.32 \\
				&\textbf{SA}  	&5723.19 &288.88	&6476.33&5122.58 \\
				&\textbf{TS}  	&8944.10	&321.54	&9522.97 &8287.20 \\
				&\textbf {PSOD2P}  &\textbf{4608.16}	&67.07	&4733.52 &4427.93 \\\hline
				\multirow{6}{*}{\textbf {Case 2}}
				&\textbf{ABC} 	&69275.63&818.78	&70452.06 &67569.59 \\
				&\textbf{GA} 	&55795.94 &1131.89	&58408.71&53144.74 \\
				&\textbf{DE}  	&85934.94&960.91	&88387.79 &83188.24 \\
				&\textbf{SA}  	&28090.84&1161.00	&31100.04 &25921.37 \\
				&\textbf{TS}  	&50733.53&1085.51	&53364.74 &48715.45 \\
				&\textbf {PSOD2P}  &\textbf{10758.48}	&124.50	&10917.75&10427.79 \\\hline
				\multirow{6}{*}{\textbf {Case 3}}
				&\textbf{ABC} 	&136892.35&1235.45	&138836.39&132516.02 \\
				&\textbf{GA}   	&112420.97&2069.67	&116193.32&107805.58 \\
				&\textbf{DE}   	&157734.59&1407.61	&160045.53&154587.73 \\
				&\textbf{SA}  	&56525.65&1036.81	&58599.25&54665.08\\
				&\textbf{TS}  	&106623.92&1638.90	&109542.78&103468.69 \\
				&\textbf {PSOD2P}  &\textbf{15317.24}	&116.32	&15557.45 &15116.52 \\\hline
				
		\end{tabular}}
		\label{numerical results of CTOP}
	\end{center}
\end{table}

\begin{table}[h]
	\begin{center}
		\caption{Numerical statistical results of running times (s) obtained by PSOD2P and other comparison algorithms for solving CTOP}
		\setlength{\tabcolsep}{2.2mm}
		{\begin{tabular}{l|l|l|l|l|l}\hline
				
				& &{Mean} &{Std.} &{Maximum} &{Minimum} \\
				\hline
				\multirow{6}{*}{\textbf {Case 1}}
				&\textbf{ABC} 	&1.58 &0.11	&2.05 &1.50 \\
				&\textbf{GA}  	&\textbf{0.19} 	&0.08 	&0.56  &0.13  \\
				&\textbf{DE}  	&0.48  &0.05 	&0.68& 0.42 \\
				&\textbf{SA}  	&8.35 &0.05	&8.47 &8.27  \\
				&\textbf{TS}  	&0.60 	&0.12 	& 1.20 & 0.53 \\
				&\textbf {PSOD2P}  &3.83 	&0.12 	& 4.25 & 3.60\\\hline
				\multirow{6}{*}{\textbf {Case 2}}
				&\textbf{ABC} 	&56.62 &0.31	&57.34 & 56.00 \\
				&\textbf{GA} 	&\textbf{0.81}  & 0.16	&1.61 &0.72  \\
				&\textbf{DE}  	&1.36 &0.04	&1.48  &1.29  \\
				&\textbf{SA}  	&14.39 &0.06	&14.54 &14.32 \\
				&\textbf {PSOD2P}  &117.65	&1.37	&122.50 &115.83 \\\hline
				\multirow{6}{*}{\textbf {Case 3}}
				&\textbf{ABC} 	&262.25 & 4.72	&281.11 &258.89  \\
				&\textbf{GA}   	&1.90 & 0.27	&3.28  & 1.75 \\
				&\textbf{DE}   	&\textbf{1.89} & 0.09	&2.12 & 1.68\\
				&\textbf{SA}  	&20.65 &0.09 	&20.94 & 20.48\\
				&\textbf{TS}  	&3.79 & 0.33	&4.58 & 3.48 \\
				&\textbf {PSOD2P}  &516.58 	&1.90	&521.36 &512.58 \\\hline		
		\end{tabular}}
		\label{CPU time results of CTOP}
	\end{center}
\end{table}

\par
\begin{figure*}[htb]
	\centering{\includegraphics[width=7in]{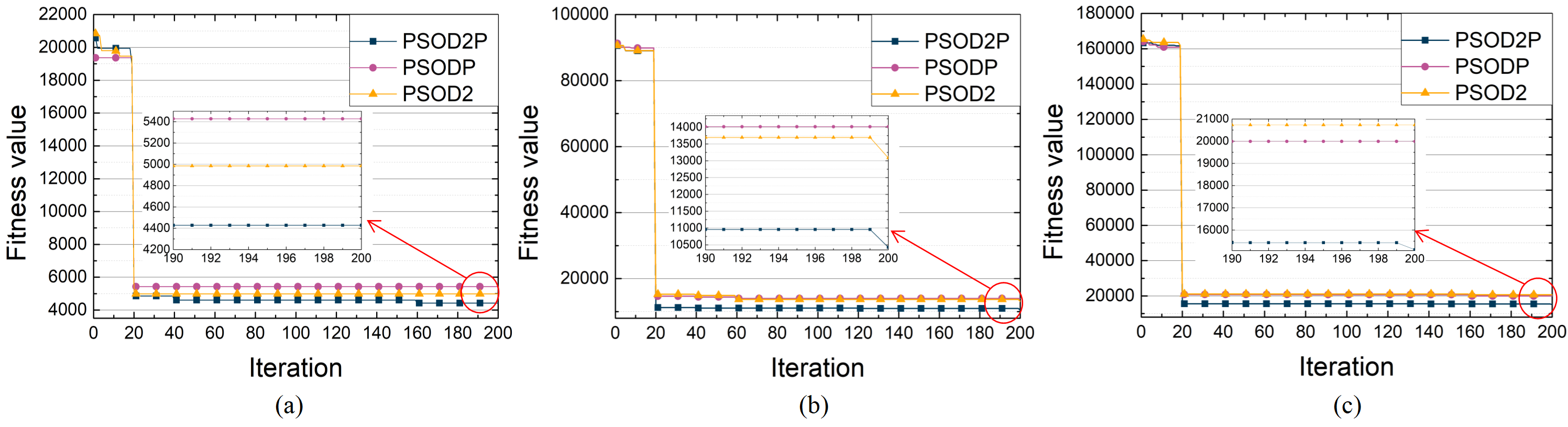}}
	\caption{Effectiveness verification for the improved factors of PSOD2P in different cases. (a) Case 1. (b) Case 2. (c) Case 3.}
	\label{IMPROVERD}
\end{figure*}

 We further conduct tests to verify the effectiveness of the introduced improved factors of PSOD2P. Since CTOP is a discrete problem, the discretization improvement factor is necessary for this problem. Thus, in this test, the optimization results, i.e., the total flight distances of CUAV obtained by PSO with a discretization improvement factor and a 2-opt operator (PSOD2), PSO with a discretization improvement factor and a path crossover reduction mechanism (PSODP) and PSOD2P are presented, respectively. Note that these tests are also independently run for 30 times in each case to prevent the random bias. Fig. \ref{IMPROVERD}(a), \ref{IMPROVERD}(b) and \ref{IMPROVERD}(c) show the corresponding test results for cases 1, 2 and 3, respectively. As can be seen, compared to conventional PSO, the discretization improvement factor provides a prerequisite for the solution of CTOP, and the 2-opt operator improves the convergence rate of the algorithm. Moreover, the path crossover reduction mechanism also helps to improve the convergence rate of the algorithm. Accordingly, the combination of these three improved factors can improve the performance of the algorithm, and the reasons are as follows. The generation method of new solutions in traditional methods is random, which may cause some newly generated solutions are worse than the original ones due to the extensive solution space, and the proposed path crossover reduction mechanism can provide a reliable method to enhance the quality of the new solutions. Specifically, there are two cases that need to be considered. First, if the original path does not have any crossover, the quality of the new solution is the same as that of the original one. Second, if the original path contains at least one crossover, the quality of the new solution is better than the original one. However, this mechanism may fall into premature convergence. Thus, the proposed 2-opt operator is able to make the algorithm to jump out from local optima. Accordingly, these two operators may make PSOD2P to be superior for deal with CTOP.

%
%
\subsection{Stability tests of the proposed algorithms}

\par
\begin{figure*}[htb]
	\centering{\includegraphics[width=7in]{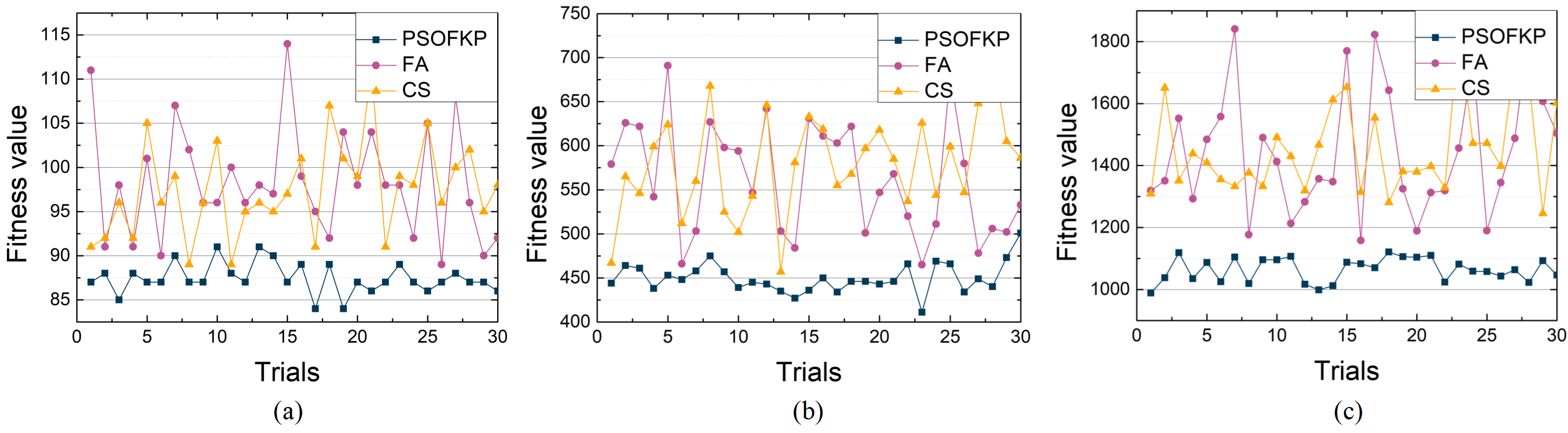}}
	\caption{The stability of different algorithms for solving CSOP. (a) Case 1. (b) Case 2. (c) Case 3.}
	\label{CSOP-stabilist}
\end{figure*}

\begin{figure*}[htb]
	\centering{\includegraphics[width=7in]{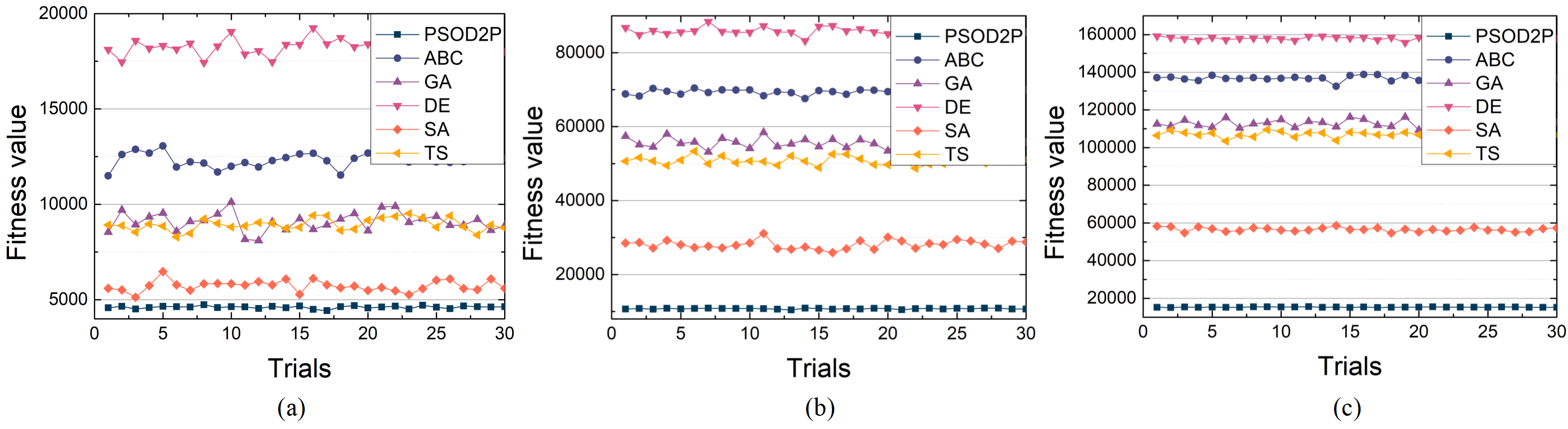}}
	\caption{Stability tests of different algorithms for solving CTOP in different cases. (a) Case 1. (b) Case 2. (c) Case 3.}
	\label{CTOP-stabilist}
\end{figure*}

 In this section, trials are conducted to evaluate the stability of the proposed algorithm. Specifically, we use PSOFKP and PSOD2P and other comparison algorithms to solve the converted CSOP and CTOP independently for 30 times, respectively, and the results are shown in Figs.~\ref{CSOP-stabilist} and \ref{CTOP-stabilist}. As can be seen, both of the proposed two algorithms can achieve the overall best stability performance on solving the corresponding optimization problems, and this can be also reflected in Tables \ref{numerical results of CSOP} and \ref{numerical results of CTOP}, respectively.

\section{Conclusion}
\label{Conclusion}

\par In this paper, the joint scheduling and trajectory optimization of CUAV for improving the charging efficiency is investigated. First, we consider to use a CUAV to charge all SNs in WRSN. Specifically, we formulate a JSTOP to jointly minimize the hovering points of CUAV, the number of repeatedly covered SNs and the flying distance of CUAV for charging all SNs. Then, due to the NP-hardness and complex hybrid solution space of JSTOP, we divided this problem into CSOP and CTOP, respectively, and propose a PSOFKP algorithm and a PSOD2P algorithm to solve the converted sub-problems. Simulation results demonstrate the superiority of the proposed joint optimization approach. Specifically, PSOFKP and PSOD2P have the overall best performance in comparison with some other benchmark algorithms, and the effectiveness of the introduced improved factors are evaluated. Moreover, it is demonstrated that both of the proposed two PSO-based algorithms are stable for solving the converted optimization problems. The results of this paper can be further extended by considering more network structure and different CUAV flight altitude, which will be investigated as the future work.

{\color{black}
\bibliographystyle{IEEEtran}
\bibliography{myref}

}

\end{document}